\documentclass[10pt,journal,compsoc]{IEEEtran}
\usepackage[table,xcdraw]{xcolor}
\usepackage{booktabs}
\IEEEoverridecommandlockouts
\usepackage[numbers]{natbib}
\usepackage{xcolor}
\usepackage{graphicx}
\usepackage{colortbl}
\usepackage{multirow}
\usepackage{tabularx}
\usepackage{flushend}
\usepackage{hyphenat}

\ifCLASSINFOpdf
\else
\fi
\AtBeginDocument{%
  \providecommand\BibTeX{{%
    \normalfont B\kern-0.5em{\scshape i\kern-0.25em b}\kern-0.8em\TeX}}}

\newif\ifdraft
\draftfalse
\def\boldification #1 {\ifdraft\textbf{#1\newline\indent}\else\relax\fi}

\newcommand{\MyBox}[1]{\vspace{3mm}\noindent\framebox[\columnwidth][c]{\parbox[b]{0.95\columnwidth}{ #1 }}\vspace{3mm}}

\begin{document}

\title{Pots of Gold at the End of the Rainbow:\\ What is Success for Open Source Contributors?}
%\title{The Pot of Gold at the End of the Rainbow:\\ What is Success for Open Source Contributors?}
%\title{What is Success for Open Source Contributors? \\ A Qualitative Study}

 \author{Bianca Trinkenreich,
        Mariam Guizani,
        Igor Wiese,
        Tayana Conte,
        Marco Gerosa,
        Anita Sarma,
        Igor Steinmacher% <-this % stops a space
\IEEEcompsocitemizethanks{\IEEEcompsocthanksitem Bianca Trinkenreich and Marco Gerosa are with the Northern Arizona University, USA.\protect
\IEEEcompsocthanksitem Mariam Guizani and Anita Sarma are with School of Electrical Engineering and Computer Science of the Oregon State University, USA. \protect
\IEEEcompsocthanksitem Igor Wiese and Igor Steinmacher are with the Department of Computer Science, Federal University of Technology – Paraná (UTFPR), Brazil.\protect
\IEEEcompsocthanksitem Tayana Conte is with the Institute of Computing (IComp) of the Federal University of Amazonas (UFAM), Manaus, AM, Brazil.}% 
}

\IEEEtitleabstractindextext{
\begin{abstract}
Success in Open Source Software (OSS) is often perceived as an exclusively code-centric endeavor. This perception can exclude a variety of individuals with a diverse set of skills and backgrounds, in turn helping exacerbate the current diversity \& inclusion imbalance in OSS. Because one's perspective of success can affect one's personal, professional, and life choices, to support a diverse class of individuals we must first understand how OSS contributors understand success. Thus far, research has used a uni-dimensional, code-centric lens to define success. In this paper, we challenge this status quo to reveal OSS contributors' multifaceted definitions of success. We do so through interviews with 27 OSS contributors whose communities recognize them as successful, and a follow-up open survey with 193 OSS contributors. Our study provides nuanced definitions of success perceptions in OSS, which might help devise strategies to attract and retain a diverse set of contributors, helping them attain their unique ``pot of gold at the end of the rainbow".
\end{abstract}

 }

\maketitle

\markboth{IEEE Transactions on Software Engineering}{}

\IEEEpeerreviewmaketitle

\section{Introduction}

\IEEEPARstart{S}uccess in Open Source Software (OSS) encompasses much more than code contributions.
%; In the words of one of our participants: Success in OSS is \textit{``contributing more than code, [and involves] contributing documentation, processes \ldots the governance of the project''}.
%as evident from one of our participant's OSS journey, who said: it is \textit{``contributing more than code, [and involves] contributing documentation, processes \ldots the governance of the project''} (P7).
However, there is a prevailing misperception that programming skills determine success in OSS~\cite{trinkenreich2020pathways}.
%However, currently there is a misperception that being successful in OSS requires ``hacking'' away at software, and that all activities in OSS are related to source code~\cite{lakhani2003hackers, fitzgerald2006transformation, robles2019twenty, Steinmacher.Teenager:2017}.
This perception is apparent in how projects highlight programming-related metrics on their sites (e.g., number of lines of code, number of commits) and in how they determine advancement in roles (e.g., what it takes to become a core member/maintainer or gain commit access)---all code-centric concepts~\cite{almarzouq2015, colazo2014performance}. Academic research has also, perhaps inadvertently, added to this misperception, as past studies have largely been code-focused. For instance, numerous papers recognize the ``onion model'' as \textit{the} mechanism through which contributors join, grow, and receive \textit{commit access} to the code repository~\cite{jergensen2011onion, agrawal2018we, wang2020unveiling}.

OSS contributors, however, are a heterogeneous group, with differing talents, skills, career goals, and motivations~\cite{harsworking, hertel2003motivation, von2012carrots, ghosh2005understanding}. Some perform a variety of non-code related activities (e.g., advocacy, technical writing, translation, project management)~\cite{trinkenreich2020pathways,carillo2017makes} and follow a different pathway than the acclaimed ``onion model''~\cite{nafus2012patches, trainer2015personal, trinkenreich2020pathways}. Given that OSS communities involve many more players than simply their ``code warriors", success must be recognized as entailing more than just the quantity of code one produces. 

%The perceptions of success help organizations to predict their employees’ commitment \cite{visagie2014means}. Imagined future affects human behavior \cite{frank1938time,lewin1936dynamic}. Motivation is also a future oriented concept \cite{francca2018motivation} that includes factors that energize and sustain human behaviour over time \cite{locke1990work}. While having perspectives of future success can enhance people’s impetus to achieve it \cite{vasquez2007seeing}, the perceptions of success and motivations constitute two different constructs.}
%In the context of OSS, the literature about motivations focuses on the joining process rather than motivation to stay active \cite{gerosa2021motivation,harsworking,hertel2003motivation,ghosh2002free,roberts2006understanding}.

How people define success impacts the choices they make in their personal and professional lives and how they evaluate others. Definitions of success can affect educational choices, decisions about where to work, project involvement, career attainment, life satisfaction, and so on~\cite{dyke2006we}. 
Thus far, the OSS literature has largely investigated the immediate motivation to join and continue in OSS~\cite{harsworking, hertel2003motivation, von2012carrots, ghosh2005understanding}, overlooking contributors' perceptions of success. Perceptions of success represent long-term goals and an imagined career future~\cite{frank1938time,lewin1936dynamic}, which influences commitment~\cite{visagie2014means} and human behavior~\cite{frank1938time,lewin1936dynamic}.
Therefore, it is important to comprehensively understand the multitude of factors that underlie what success means to an individual in OSS. Without such an understanding, how can OSS communities support the many diverse individuals whose future goals and pathways do not fit the typical onion model career mold? 

In this study, we tackle the fundamental research question:\textit{ What does it mean to be successful in OSS?}
%\begin{itemize}
%\item[]\textbf{RQ.} What does it mean to be successful in OSS?
%\end{itemize}

To answer this question, we interviewed 27 OSS contributors who are recognized as successful figures in their communities. We qualitatively analyzed the interviews using the ``success model" proposed by Dries et al.~\cite{dries2008career}. We then triangulated our results with data from a survey of 193 OSS contributors.

Our results indicate that OSS contributors define success in multi-faceted and nuanced ways. Success includes both objective measures (e.g., monetary compensations, amounts of contribution) as well as subjective ones (e.g., recognition in the community, satisfaction). 
%And while the end goal of contributors may be similar, how they go about being successful can be very different for different people and is rooted in their perceptions of what success means to them. 
In the words of von~Krogh~\cite{von2012carrots}: ``\textit{Occasionally, humans also make elaborate detours, strive for bigger things in life, and undertake long voyages to find the gold at the end of the rainbow}.'' Thus, it is time that we reflect on what we consider success in OSS, and how we can help make OSS more diverse by finding different ways to support individuals with various backgrounds and who have distinct definitions of success.

In this paper, we introduce the definition of career success and the theoretical success model we use in Section~\ref{sec:background}, followed by our research method and results in Sections~\ref{sec:research_design} and~\ref{sec:results}, respectively. Sections~\ref{sec:discussion} \& \ref{sec:recommendations} discuss the implications of our results, followed by related work, limitations, and conclusions in Sections~\ref{sec:relatedWork}, ~\ref{sec:limitations}, and Section~\ref{sec:conclusion}.

%This is a call to action to our fellow researchers: join us in helping make OSS diverse by finding different ways to support diverse individuals with diverse backgrounds and motivations and who have diverse definitions of success.

%\review{The paper is organized as follows. Section~\ref{sec:background} discusses the definition of career success. Sections~\ref{sec:research_design} and~\ref{sec:results} present the research method and results, respectively. Sections~\ref{sec:discussion} and \ref{sec:recommendations} discuss the results and implications for practitioners and researchers. Finally, Section~\ref{sec:relatedWork} presn, limitations in Section~\ref{sec:limitations}, and conclusions in Section~\ref{sec:conclusion}.}

\section{Defining Success}
\label{sec:background}

Success can be defined as ``the accomplishment of desirable work-related outcomes at any point in a person’s work experiences over time''~\cite{arthur2005career} and is a dynamic concept~\cite{savickas2005theory}. Existing literature shows that career success can be characterized from different perspectives, such as: job, interpersonal, financial stability, and life success~\cite{gattiker1986subjective}; balance, relationships, recognition, and material success~\cite{dyke2006we}; psychological, and social success~\cite{hennequin2007career}; and extrinsic and intrinsic success~\cite{sturges1999means}.

%By analyzing the literature related to career success, we chose to use the 
%, because it is not tied to organizations, offers different lenses and understandings of success, and generalizes to different contexts.
%For example, Valk et al. \cite{valk2014international} used the Dries model to explain the career success of women in Science and Technology, Benson et al. \cite{benson2020cultural} used the model to examine the effect of cultural values on career success, and Santos and Martins \cite{gaio2019linking} used to understand the reasons behind turnover between employees who are repatriates.
%The dimensions proposed on the model can be adapted not only to success linked to an organization but also to one's personal evolution regardless of the organization, both from the perspective of the one's relationships with other people and from the more intrinsic side. 
%in OSS context because, even though contributors do not follow a linear pathway, they have the experiences that constitute a boundaryless career~\cite{arthur2005career} in which a career is built upon experiences, instead of jobs.

\begin{figure}[hbt]
     \centering
     \includegraphics[width=0.42\textwidth]{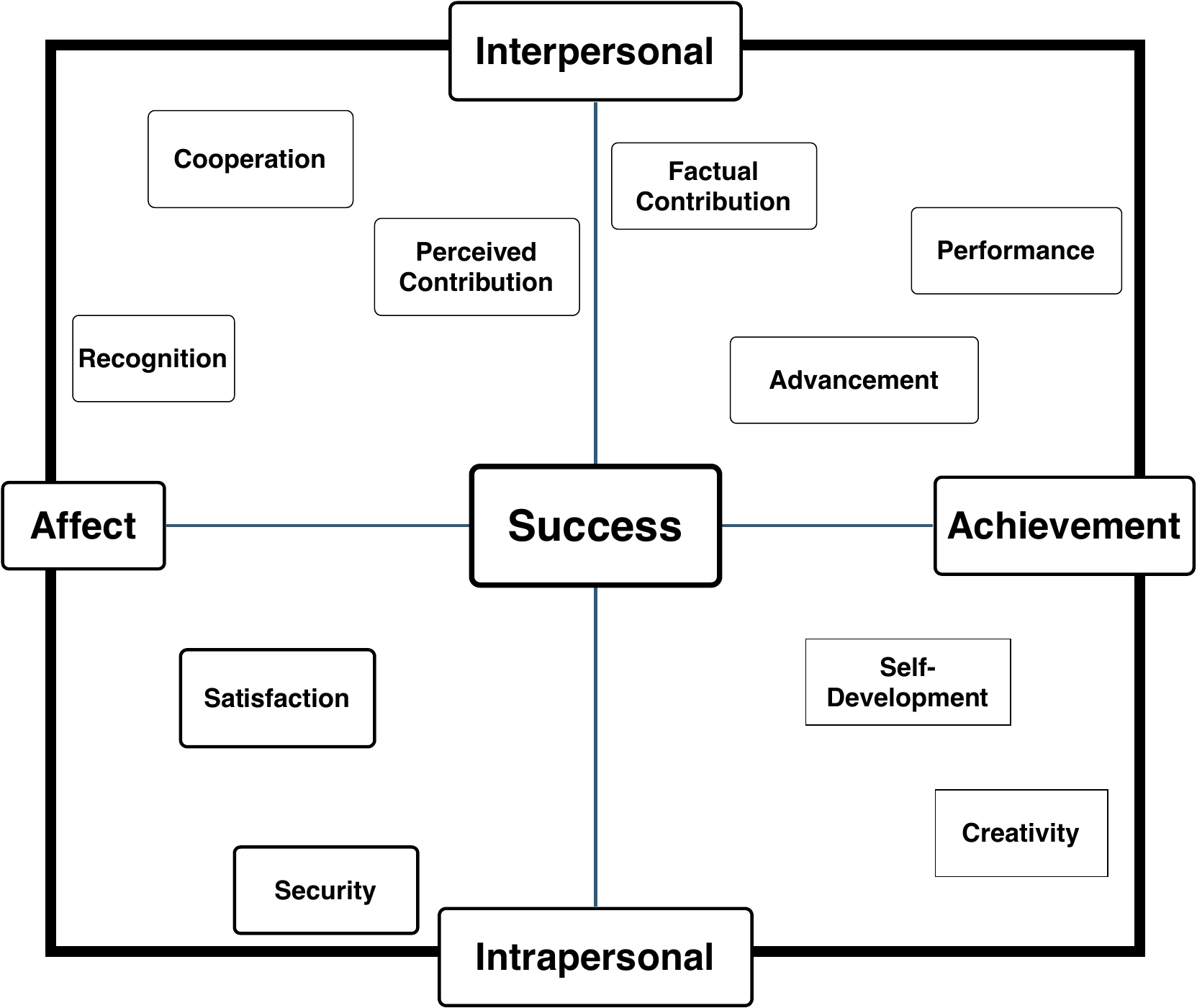}
     \vspace{-0.25cm}
     \caption{The multidimensional model of success \cite{dries2008career}}
     \vspace{-0.25cm}
     \label{fig:dries_model}
 \end{figure}

Dries et al.~\cite{dries2008career} organized existing literature enhanced with additional data collection into a comprehensive multi-dimensional theoretical model that describes how people perceive career success (Figure~\ref{fig:dries_model}). Their model has been used in multiple contexts, such as the success of women in Science and Technology~\cite{valk2014international}, the effects of cultural values on career success~\cite{benson2020cultural}, and the reasons behind turnover between employees who are repatriates~\cite{gaio2019linking}. We use this model to organize our findings since it provides a comprehensive view of success, consolidates previous literature, is not tied to a specific type of organization, and generalizes to different contexts, fitting well to the heterogeneity of OSS and the boundaryless career path that OSS contributors follow.

%The chosen model~\cite{dries2008career} is a comprehensive multidimensional theoretical model to explain success and encompasses several concepts from the other studies~\cite{gattiker1986subjective,schein1990career,sturges1999means,dyke2006we,lee2006exploring,hennequin2007career,parker1991motivation,nabi2001relationship,hall1976careers,hall1979career} . 

Dries et al.~\cite{dries2008career}'s model comprises two dimensions, further classified into four quadrants and ten regions. The first dimension is \textit{Affect $\times$ Achievement}. \textit{Affect} represents the subjective feelings and perceptions that cause people to weigh their success as high or low. \textit{Achievement} represents the objective side: the factual accomplishments through which people measure their success. 

The second dimension is \textit{Interpersonal $\times$ Intrapersonal}. \textit{Interpersonal} involves one's relationships and interactions with the outside world. \textit{Intrapersonal} indicates one's ``self'': their internal world. The combination of these two dimensions (Affect $\times$ Achievement and Interpersonal $\times$ Intrapersonal) generates four quadrants: (Quad1) Interpersonal $\times$ Affect; (Quad2) Interpersonal $\times$ Achievement; (Quad3) Intrapersonal $\times$ Achievement; (Quad4) Intrapersonal $\times$ Affect. The multidimensional nature of this model shows how success can have several different---yet complementary---meanings that may serve people with different goals.

%The Dries et al. \cite{dries2008career} model of career success is being used to contextualize success by researchers in different domains. Benson et al. \cite{benson2020cultural} examined the direct effect of cultural values on definitions of Dries model of career success while controlling for individual, professional, and national economic development differences. Santos e Martins \cite{gaio2019linking} investigated the reasons that can impact turnover intention of repatriates and used the Dries model of career success to categorize their findings.

Akkermans and Kubasch~\cite{akkermans2017trending} explain that careers have been changing over the past few decades, evolving into more complex and unpredictable endeavors that require empirical studies in different domains to understand success. We use the above model to analyze the definition of success in OSS.

\section{Research Design}
\label{sec:research_design}
\begin{figure*}[hbt]
     \centering
     \includegraphics[width=1\textwidth]{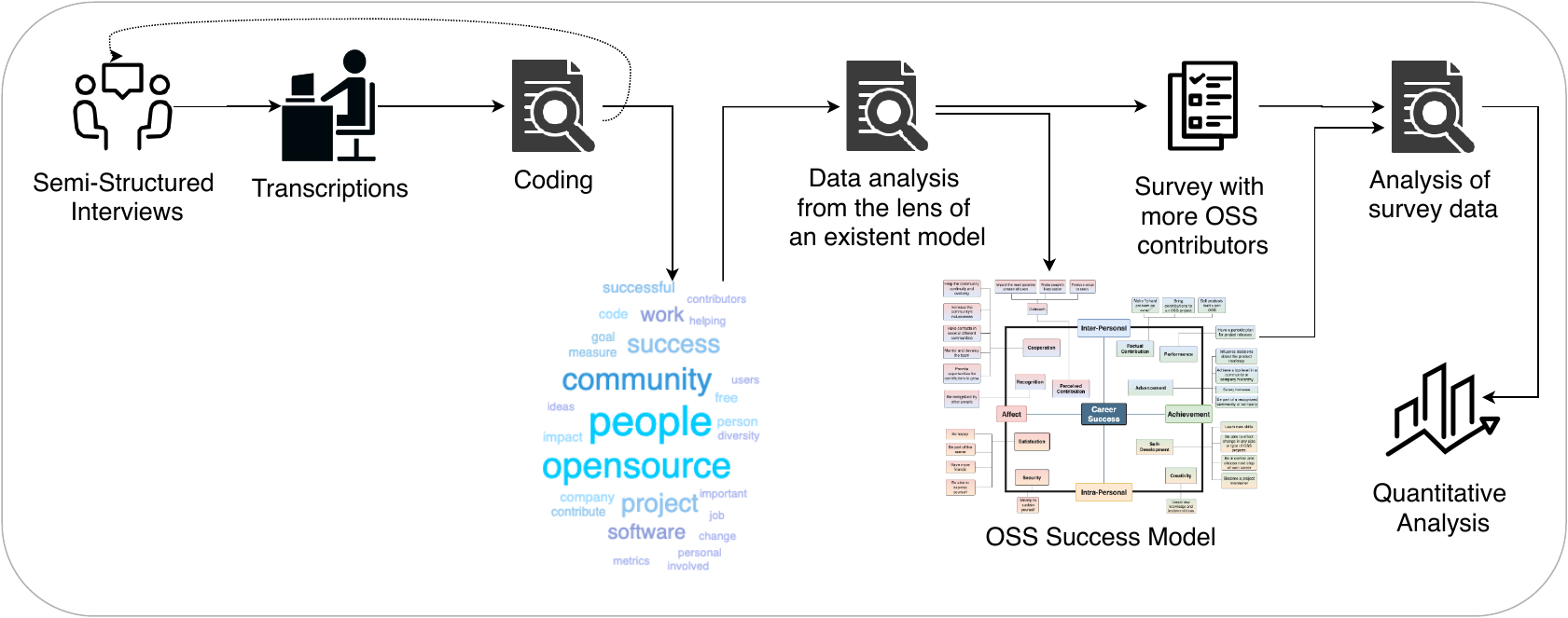}
     \caption{The research method, which included face-to-face interviews at the OSCON'19 event and later with OSS maintainers through video conferencing, as well as a large-scale survey. We conducted qualitative analysis to build the OSS Success Model and qualitative and quantitative analysis to triangulate the definition of success we found from the interviews’ data}
     \label{fig:method}
 \end{figure*}

This section presents the design of our study, which included interviews and a survey~\footnote{The research protocol was approved by the Oregon State University institutional review board (IRB).}, as depicted in Figure~\ref{fig:method}.

\subsection{Interviews: Building the OSS Success Model}
\label{phase1}

Due to the complexity of the phenomenon under study, we started with in-depth interviews to understand how OSS contributors perceive success.

\subsubsection{Interview Planning}
\label{sec:planning}
For the interviews, we aimed to recruit recognized OSS contributors to understand successful OSS career stories. We started by recruiting invited speakers of the Open Source Software Conference (OSCON-2019), a well-recognized open source conference focused on practitioners. These speakers were invited to give talks in the main lineup of the conference, suggesting they are successful in OSS. 

Before OSCON started, we emailed and sent direct messages via Twitter to all speakers whose contact information was publicly available (15). We also approached some of them during the event. During the event, we conducted 11 face-to-face interviews. In addition, we used a snowball approach to recruit more interviewees. At the end of each interview we asked them to introduce us to other qualified participants for the study. We conducted 4 additional interviews from this snowball approach. A majority of our interviewees were women, possibly because they cared about our goal. This initial imbalance was counterbalanced in the subsequent interviews (see Table~\ref{tab:demo}).

After this first cycle of interviews and analysis, we recruited 12 additional participants. We invited maintainers of mature OSS projects who could share their perspectives of contributing to OSS. In both phases of interviews (at OSCON and post event), we used a snowball approach: at the end of each interview, we asked interviewees to introduce us to other qualified participants for the study, aiming to reach other speakers and successful project maintainers. Our goal was to interview experienced contributors working at least 5 years in well-known and mature OSS projects. We compensated interviewees with a 25-dollar gift card. 

% For those who had been referred by other participants (snowballing), we checked their profiles on GitHub, LinkedIn, and personal website to validate if the person met our inclusion criteria.

Before interviewing participants, we conducted five pilot interviews with a professor and four PhD students who were experienced in OSS. The goal was to solicit feedback on the script and ensure that the interview would fit in a 40- to 60-minute time slot. We analyzed the pilot interview responses to ensure that we answered our research question with an adequate level of detail. 

% \usepackage[normalem]{ulem}
\begin{table*}[htb]
\centering
\caption{Interview Script and Survey Questions (excluding demographic questions)}
\label{tab:script}
\footnotesize
\begin{tabular}{l}
\hline
\multicolumn{1}{c}{\cellcolor{gray!35}\textbf{Interview Script}}                                                                                                               \\ \hline\hline
\begin{tabular}[c]{@{}l@{}}I-1. Can you please tell me all the story of your career? All your professional journey, from the beginning of your career until where you are today\end{tabular} \\ \hline
I-2. How would you define yourself as being a successful OSS contributor? \\ \hline
\begin{tabular}[c]{@{}l@{}}I-3. Is there any kind of success that you didn't achieve yet? What do you plan to achieve in the future?\end{tabular}     \\ \hline
\begin{tabular}[c]{@{}l@{}}I-4. Please think about a person you consider successful in OSS. Why do you think this person is successful?\end{tabular} \\ \hline
\begin{tabular}[c]{@{}l@{}}I-5. Now the opposite. Please think about a person you consider not successful in OSS. Why do you think this person is not successful?\end{tabular} \\ \hline
\multicolumn{1}{c}{\cellcolor{gray!35}\textbf{Survey Questions}}                                                                                                                   \\ \hline\hline
S-1. Do you consider yourself a successful OSS contributor? (Yes/No/I don't Know)                                                                       \\ \hline
S-2. How would you define a successful person in OSS? (Open Question)                                                                                      \\ \hline
\end{tabular}
\end{table*}

We conducted semi-structured interviews~\cite{seaman1999qualitative}. 
We used a script as we present in Table \ref{tab:script} to guide the different areas of inquiry, while also listening for unanticipated information during the flow of the conversation. 
The interviews revolved around the central question: \textit{``How would you define being successful in Open Source?''} We approached this topic after establishing rapport with the interviewee, asking about their career story and contributions. 

We interviewed participants until we could not find any new concept related to our research question for five consecutive interviews. Our final sample comprised 27 participants. Table~\ref{tab:demo} presents their demographics.

%Our initial goal was to interview at least 10 people, which would be in line with the guidelines provided by the anthropology literature---which mentions that a set of 10--20 knowledgeable people is enough to uncover and understand the core categories in a study of lived experience~\cite{bernard2017research}. However, we planned to continued recruiting and interviewing until we could not find new understandings related to success in OSS for at least five consecutive interviews. 

\begin{table*}[htb]
\centering
\caption{Demographics for the Interview Participants}
\vspace{-0.3cm}
\label{tab:demo}
\resizebox{0.83\textwidth}{!}{%
\begin{tabular}{c|l|r|l|l|l}
\hline
\cellcolor{gray!35}\textbf{Participant ID} & \cellcolor{gray!35}\textbf{Gender}        & \cellcolor{gray!35}\textbf{Years in OSS} & \cellcolor{gray!35}\textbf{Main Type of Contribution}  & \cellcolor{gray!35}\textbf{Recruitment} & \cellcolor{gray!35}\textbf{Interview Mode} \\ \hline
P1  & Woman                  & 6 ~~~~~~ & OSS Advocate            & OSCON Speaker & In-person      \\ 
P2  & Woman                  & 5  ~~~~~~ & OSS Coder               & Mature project   & Video-conference \\ 
P3  & Woman                  & 13 ~~~~~~ & OSS Treasurer           & OSCON Speaker & In-person       \\ 
P4  & Man                    & 9  ~~~~~~ & OSS System Admin        & OSCON Speaker & In-person       \\ 
P5  & Prefer not to say & 7  ~~~~~~ & OSS Coder               & OSCON Speaker & In-person       \\ 
P6  & Man                    & 5  ~~~~~~ & OSS Coder               & OSCON Speaker & In-person       \\ 
P7  & Man                    & 12 ~~~~~~ & OSS Coder               & OSCON Speaker & In-person       \\ 
P8  & Woman                  & 30 ~~~~~~ & OSS Strategist          & OSCON Speaker & In-person       \\ 
P9  & Man                    & 13 ~~~~~~ & OSS Advocate            & OSCON Speaker & In-person       \\ 
P10 & Woman                  & 20 ~~~~~~ & OSS Advocate            & OSCON Speaker & In-person       \\ 
P11 & Woman                  & 20 ~~~~~~ & OSS Writer              & Snowballing      & Video-conference \\ 
P12 & Woman                  & 20 ~~~~~~ & OSS Advocate and Writer & OSCON Speaker & In-person       \\ 
P13 & Woman                  & 7  ~~~~~~ & OSS Advocate            & Mature project   & Video-conference \\ 
P14 & Woman                  & 20 ~~~~~~ & OSS License Manager     & Mature project   & Video-conference \\ 
P15 & Woman                  & 15 ~~~~~~ & OSS Advocate            & OSCON Speaker    & In-person       \\ 
P16 & Woman                  & 10 ~~~~~~ & OSS Advocate            & Snowballing      & Video conference \\ 
P17 & Woman                  & 5  ~~~~~~ & OSS Project Manager     & Snowballing & Video conference       \\ 
P18 & Man                    & 8  ~~~~~~ & OSS Coder               & Mature project   & Video-conference \\ 
P19 & Man                    & 8  ~~~~~~ & OSS Coder               & Mature project   & Video-conference \\ 
P20 & Man                    & 5  ~~~~~~ & OSS Coder               & Mature project   & Video-conference \\ 
P21 & Man                    & 15 ~~~~~~ & OSS Coder               & Mature project   & Video-conference \\ 
P22 & Man                    & 10 ~~~~~~ & OSS Advocate            & Mature project   & Video-conference \\ 
P23 & Man                    & 7  ~~~~~~ & OSS Coder               & Snowballing      & Video-conference \\ 
P24 & Man                    & 20 ~~~~~~ & OSS Coder               & Mature project   & Video-conference \\ 
P25 & Man                    & 23 ~~~~~~ & OSS Coder               & Mature project   & Video-conference \\ 
P26                     & Prefer not to say & 10                    ~~~~~~ & OSS Project Manager & Mature project       & Videoconference         \\ 
P27 & Woman                  & 10 ~~~~~~ & OSS Coder               & Mature project   & Video-conference \\ \hline
\end{tabular}
}
\end{table*}

\subsubsection{Data Collection}
\label{sec:collection}
Five researchers participated in conducting the interviews, where there were at least two researchers per interview. The researchers have at least six years' experience in qualitative studies. The interviews were face-to-face during OSCON and over video conference calls afterward. Interviews lasted between 40 and 60 minutes. With participant consent, we recorded all interviews. The first author of this paper transcribed the interviews using \textsc{Otter.ai}\footnote{https://otter.ai} and listened to each recording, adjusting the corresponding transcriptions, mainly regarding technical terms and project names.

% For those interviewed online, we arranged a convenient meeting time and sent out a consent letter via e-mail. 

Our sample comprises paid and volunteer contributors across 20 different OSS projects (e.g., Kubernetes, Drupal, R, Noosfero, Fedora, Debian, GitLab), which vary in terms of number of contributors (30 to 3,000 contributors), product domains (including infrastructure and user-application projects), and types (backed by foundations, communities, and companies). Table~\ref{tab:demo} presents the demographics of our sample. Because of the terms of consent, we cannot link each participant to their project.

\subsubsection{Data Analysis}
\label{sec:analysis}
 
The data analysis was performed in two stages. In the first stage we analyzed the interview data collected at OSCON 2019 and in the second stage we analyzed the data from the additional interviews. 

We qualitatively analyzed the transcripts of the interviews by inductively applying open coding in groups, wherein we identified the definition of success that each participant provided. We built post-formed codes as the analysis progressed and associated them with respective parts of the transcribed text, so as to code the success definitions according to the participants' perspectives, who were identified as P1 to P27.

The outcome was a set of higher-level categories as cataloged in our codebook~\footnote{https://figshare.com/s/39491da83e398612dffa}.

%As suggested by Strauss and Corbin~\cite{strauss1997grounded}, after performing the initial data analysis, we looked at the literature for existing models that represent the multiple facets and perspectives of success. We chose the model proposed by Dries et al.~\cite{dries2008career} (see Section~\ref{sec:background}) as it models a multi-dimensional perspective of success. 

To organize our categories according to Dries et al.'s model~\cite{dries2008career} (see Section~\ref{sec:background}), three of the authors conducted multiple card sorting sessions together~\cite{spencer2009}, arranging the codes according to the regions of the model. After the initial sorting, the group met once a week for four weeks to discuss and validate the results with the other authors. The process of categorizing the codes into the regions of Dries et al.'s model~\cite{dries2008career} was conducted using continuous comparison~\cite{Strauss.Corbin_1998} and negotiated agreement~\cite{garrison2006revisiting} (as a group). In the negotiated agreement process, the researchers discussed the rationale they used to categorize each code until they reached consensus~\cite{garrison2006revisiting}.
%%%Igor removed --> forman2007qualitative
\subsection{Survey: Data Triangulation}
\label{sec:phase2}

Next, we conducted an online survey to triangulate the interview results by gathering data from a different perspective~\cite{easterbrook2008selecting} and a larger sample.

\subsubsection{Survey Planning}
\label{sec:survey_planning}
In the survey, we asked two key questions about participants' perceptions of success (see Table ~\ref{tab:script}), and additional demographic-related questions, including the relationship with OSS (paid/unpaid), types of contributions, gender identity, country of residences, and age. The target population included any person who contributes to OSS.

% a closed question (``Do you consider yourself as a successful OSS contributor?") and

%We evaluated the instrument by pre- and pilot-tests by six professors who used both computer and mobile phones. 

We advertised the survey on social media and community blogs (e.g., Linkedin, Twitter, Facebook, Reddit, Hackernews, CHAOSS blog, and others). To reach a broader audience, we paid to promote our posts on Twitter, Facebook, and Reddit. We also sent direct messages to OSS contributors and discussion lists. We offered the participants a chance to enter a raffle for US\$100 to increase the response rate. To enter the raffle, they needed to opt-in and provide an email address at the end of the survey.

\subsubsection{Data Collection}
\label{sec:survey_collection}
The survey was available between June 4$^{th}$ and July 3$^{rd}$, 2020. We received 217 non-blank responses. We filtered our data to consider only valid responses. We analyzed the attention check answers, time to complete the questionnaire, equal/similar e-mail addresses, and inappropriate answers to the open questions (e.g., ``I am the POTUS,'' ``I don't wanna answer''), resulting 193 valid responses.

%We inspected these answers and removed 24 that did not provide a relevant answer (e.g., ``I don't know", ``I don't want to answer"). The final analysis included 193 answers.

%dropped answers that failed the attention check question (0 cases), that took  lower outliers (0 removed). We manually inspected the open text question, looking for senseless and inappropriate answers (24 removed), as for example ``I don't know", ``I'm not sure", "hard to say", ``I don't want to answer", ``I am the POTUS". Then, we filtered our data looking for potential duplicate participation, even though the survey platform (Qualtrics) has mechanisms to prevent multiple responses from the same participant. We started looking for identical and similar emails (0 removed). Finally, since this study’s target population comprises OSS project contributors, we also inspected the answers to the question about years of experience in OSS to filter answers from participants with no experience (0 removed). After applying all the filters, we ended up with 193 participants.}

We asked participants their three main types of contributions and classified participants as ``coder" if they selected "code developer" or ``code reviewer'' as one of the three main types of contributions. We classified as non-coders those who selected only a subset of these options: translation, documentation, mentorship, user support, community building, bug triaging, event presentations, advocacy and evangelism, creative work and design, and project management. We present the demographics of the survey participants in Table \ref{tab:survey_demo}. 

\begin{table}[htb]
\centering
\caption{Demographics for the Survey Respondents}
\label{tab:survey_demo}
\resizebox{0.48\textwidth}{!}{%
\begin{tabular}{ccccccc}
\cline{1-3} \cline{5-7}
\multicolumn{1}{|c|}{\cellcolor[HTML]{C0C0C0}Type of Contribution} &
  \multicolumn{1}{c|}{\cellcolor[HTML]{C0C0C0}\#} &
  \multicolumn{1}{c|}{\cellcolor[HTML]{C0C0C0}\%} &
  \multicolumn{1}{c|}{} &
  \multicolumn{1}{c|}{\cellcolor[HTML]{C0C0C0}Country of Residence} &
  \multicolumn{1}{c|}{\cellcolor[HTML]{C0C0C0}\#} &
  \multicolumn{1}{c|}{\cellcolor[HTML]{C0C0C0}\%} \\ \cline{1-3} \cline{5-7} 
\multicolumn{1}{|c|}{Coder} &
  \multicolumn{1}{c|}{163} &
  \multicolumn{1}{c|}{84.46\%} &
  \multicolumn{1}{c|}{} &
  \multicolumn{1}{c|}{Germany} &
  \multicolumn{1}{c|}{89} &
  \multicolumn{1}{c|}{46.11\%} \\ \cline{1-3} \cline{5-7} 
\multicolumn{1}{|c|}{Non-Coder} &
  \multicolumn{1}{c|}{30} &
  \multicolumn{1}{c|}{15.54\%} &
  \multicolumn{1}{c|}{} &
  \multicolumn{1}{c|}{USA} &
  \multicolumn{1}{c|}{59} &
  \multicolumn{1}{c|}{30.57\%} \\ \cline{1-3} \cline{5-7} 
\multicolumn{3}{l}{} &
  \multicolumn{1}{c|}{} &
  \multicolumn{1}{c|}{Netherlands} &
  \multicolumn{1}{c|}{12} &
  \multicolumn{1}{c|}{6.22\%} \\ \cline{1-3} \cline{5-7} 
\multicolumn{1}{|c|}{\cellcolor[HTML]{C0C0C0}Gender} &
  \multicolumn{1}{c|}{\cellcolor[HTML]{C0C0C0}\#} &
  \multicolumn{1}{c|}{\cellcolor[HTML]{C0C0C0}\%} &
  \multicolumn{1}{c|}{} &
  \multicolumn{1}{c|}{Brazil} &
  \multicolumn{1}{c|}{9} &
  \multicolumn{1}{c|}{4.66\%} \\ \cline{1-3} \cline{5-7} 
\multicolumn{1}{|c|}{Men} &
  \multicolumn{1}{c|}{165} &
  \multicolumn{1}{c|}{85.49\%} &
  \multicolumn{1}{c|}{} &
  \multicolumn{1}{c|}{Others} &
  \multicolumn{1}{c|}{23} &
  \multicolumn{1}{c|}{11.92\%} \\ \cline{1-3} \cline{5-7} 
\multicolumn{1}{|c|}{Women} &
  \multicolumn{1}{c|}{16} &
  \multicolumn{1}{c|}{8.29\%} &
  \multicolumn{1}{c|}{} &
  \multicolumn{1}{c|}{Did not answer} &
  \multicolumn{1}{c|}{1} &
  \multicolumn{1}{c|}{0.52\%} \\ \cline{1-3} \cline{5-7} 
\multicolumn{1}{|c|}{Non-Binary} &
  \multicolumn{1}{c|}{2} &
  \multicolumn{1}{c|}{1.04\%} &
  \multicolumn{1}{l}{} &
  \multicolumn{3}{l}{} \\ \cline{1-3} \cline{5-7} 
\multicolumn{1}{|c|}{Did not answer} &
  \multicolumn{1}{c|}{10} &
  \multicolumn{1}{c|}{5.18\%} &
  \multicolumn{1}{c|}{} &
  \multicolumn{1}{c|}{\cellcolor[HTML]{C0C0C0}Age} &
  \multicolumn{1}{c|}{\cellcolor[HTML]{C0C0C0}\#} &
  \multicolumn{1}{c|}{\cellcolor[HTML]{C0C0C0}\%} \\ \cline{1-3} \cline{5-7} 
\multicolumn{3}{l}{} &
  \multicolumn{1}{c|}{} &
  \multicolumn{1}{c|}{24 or less} &
  \multicolumn{1}{c|}{30} &
  \multicolumn{1}{c|}{15.54\%} \\ \cline{1-3} \cline{5-7} 
\multicolumn{1}{|c|}{\cellcolor[HTML]{C0C0C0}Financial Relation} &
  \multicolumn{1}{c|}{\cellcolor[HTML]{C0C0C0}\#} &
  \multicolumn{1}{c|}{\cellcolor[HTML]{C0C0C0}\%} &
  \multicolumn{1}{c|}{} &
  \multicolumn{1}{c|}{25-34} &
  \multicolumn{1}{c|}{60} &
  \multicolumn{1}{c|}{31.09\%} \\ \cline{1-3} \cline{5-7} 
\multicolumn{1}{|c|}{Paid} &
  \multicolumn{1}{c|}{36} &
  \multicolumn{1}{c|}{18.65\%} &
  \multicolumn{1}{c|}{} &
  \multicolumn{1}{c|}{35-44} &
  \multicolumn{1}{c|}{59} &
  \multicolumn{1}{c|}{30.57\%} \\ \cline{1-3} \cline{5-7} 
\multicolumn{1}{|c|}{Unpaid} &
  \multicolumn{1}{c|}{137} &
  \multicolumn{1}{c|}{70.98\%} &
  \multicolumn{1}{c|}{} &
  \multicolumn{1}{c|}{45-54} &
  \multicolumn{1}{c|}{27} &
  \multicolumn{1}{c|}{13.99\%} \\ \cline{1-3} \cline{5-7} 
\multicolumn{1}{|c|}{Partially Paid} &
  \multicolumn{1}{c|}{16} &
  \multicolumn{1}{c|}{8.29\%} &
  \multicolumn{1}{c|}{} &
  \multicolumn{1}{c|}{55 or more} &
  \multicolumn{1}{c|}{10} &
  \multicolumn{1}{c|}{5.18\%} \\ \cline{1-3} \cline{5-7} 
\multicolumn{1}{|c|}{Did not answer} &
  \multicolumn{1}{c|}{4} &
  \multicolumn{1}{c|}{2.07\%} &
  \multicolumn{1}{c|}{} &
  \multicolumn{1}{c|}{Did not answer} &
  \multicolumn{1}{c|}{7} &
  \multicolumn{1}{c|}{3.63\%} \\ \cline{1-3} \cline{5-7} 
\multicolumn{1}{l}{} &
  \multicolumn{1}{l}{} &
  \multicolumn{1}{l}{} &
  \multicolumn{1}{l}{} &
  \multicolumn{1}{l}{} &
  \multicolumn{1}{l}{} &
  \multicolumn{1}{l}{} \\ \hline
\rowcolor[HTML]{C0C0C0} 
\multicolumn{5}{|c|}{\cellcolor[HTML]{C0C0C0}Do you consider yourself a successful OSS contributor?} &
  \multicolumn{1}{c|}{\cellcolor[HTML]{C0C0C0}\#} &
  \multicolumn{1}{c|}{\cellcolor[HTML]{C0C0C0}\%} \\ \hline
\multicolumn{5}{|c|}{Yes} &
  \multicolumn{1}{c|}{72} &
  \multicolumn{1}{c|}{37.31\%ˆ} \\ \hline
\multicolumn{5}{|c|}{No} &
  \multicolumn{1}{c|}{80} &
  \multicolumn{1}{c|}{41.45\%} \\ \hline
\multicolumn{5}{|c|}{I'm not sure} &
  \multicolumn{1}{c|}{41} &
  \multicolumn{1}{c|}{21.24\%} \\ \hline
\end{tabular}}
\end{table}

\subsubsection{Data Analysis}
\label{sec:survey_analysis}

We used the categories from the interviews, classified into the regions of Dries et al.'s model~\cite{dries2008career}, as the starting point of the qualitative analysis of the survey questions. We diligently analyzed the answers to identify any new perceptions of success that did not previously emerge from the interviews, but all survey answers could be mapped to the existing categories. We also used descriptive statistics to summarize the survey responses, their association with each other (success constructs), and the demographics data ~\cite{wohlin2015towards}. 
%We use the chi-square test at $\alpha = .05$ and Yates and Bonferroni corrections to investigate any statistical relationship between the different analysis constructs.

See supplemental material\footnote{https://figshare.com/s/39491da83e398612dffa} for additional details, including sample answers to the demographics and open question survey questions and the qualitative analysis codes.
%, as well as details of the chi-square tests.

% Survey respondents are identified as S1 to S193.

\section{Results}
\label{sec:results}

Here, we present our participants' definitions of success. 

%The survey results helped validate our understanding of success from the interviews. Additionally, we investigate how differences in demographics may affect participants' perspectives on success through a segmented analysis of the survey results in Section \ref{sec:results_segmented}.

\subsection{Understanding Success in OSS}
\label{sec:results_existing_model}

\begin{figure*}[htb]
     \centering
     \includegraphics[width=1\textwidth]{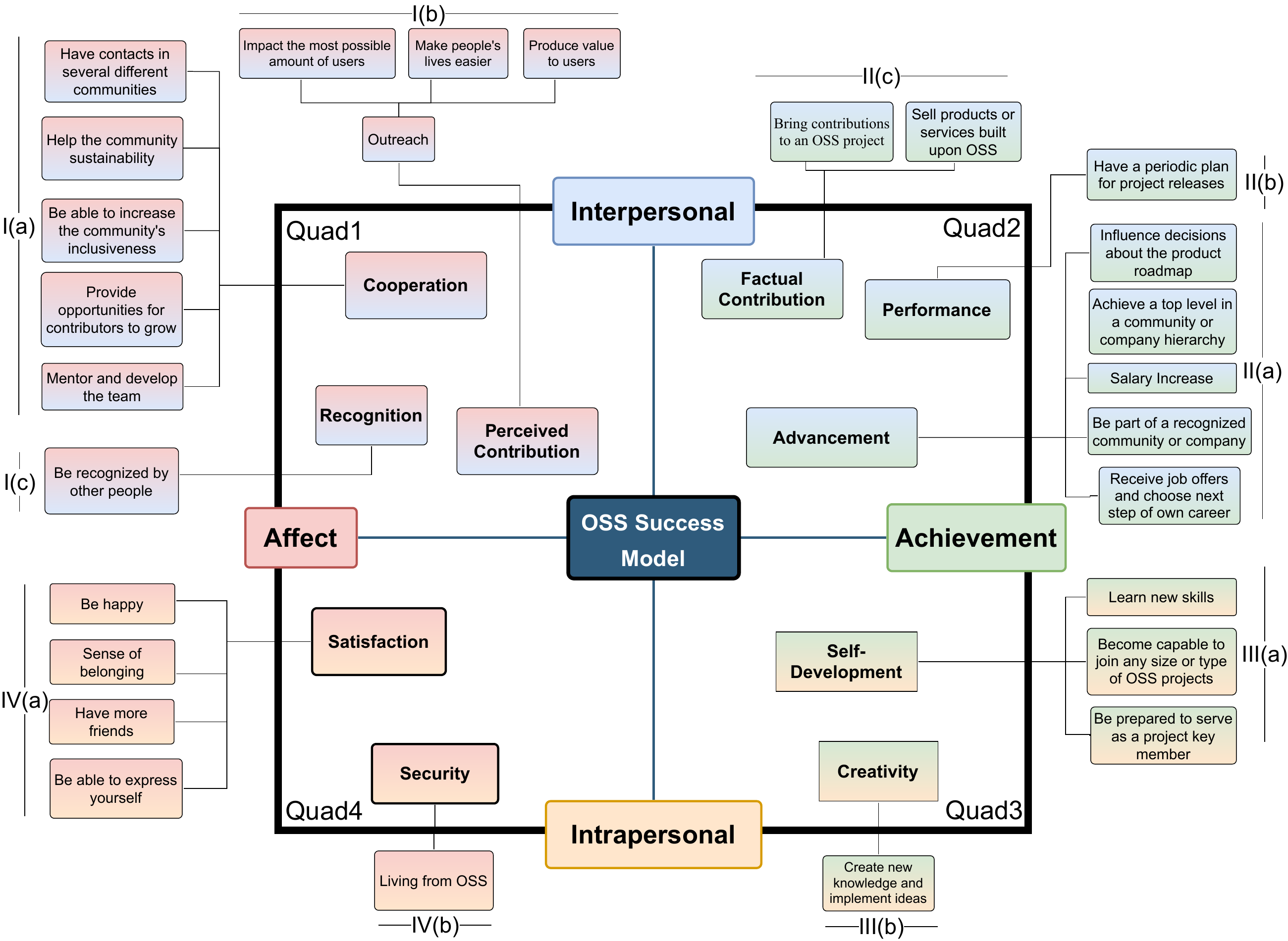}
     \caption{OSS Success Model. We mapped our participants' definitions (shown outside the bold square) to Dries et al.'s model~\cite{dries2008career}, which organize success in four quadrants.}
     \label{fig:carrer_success}
 \end{figure*}

\boldification{**We distilled the different perspectives of success from the interview data and use the Dries model.}
Our analysis of the interviews (see Section~\ref{sec:analysis}) revealed 26 categories that explain how our participants defined success. We organized these categories using the multidimensional model of success proposed by Dries et al.~\cite{dries2008career} (see Section~\ref{sec:background}), as can be seen in Figure \ref{fig:carrer_success}. The 26 categories covered all ten regions of the model. Table \ref{tab:codes} presents the number of participants (interviews and survey) whose responses fit in each region. The survey analysis did not provide any new definitions of success. In the following, we present our findings organized by quadrant. 

%Recall this model characterizes success using two dimensions: \textit{Affect x Achievement} and \textit{Inter x Intrapersonal}, and identifies \textit{nine regions} across the four quadrants defined by these dimensions. 

\begin{table*}[htb]
\centering
%\scriptsize
\caption{Success meanings from the interviews and the survey classified per Dries et al.'s  model \cite{dries2008career}}
\vspace{-0.3cm}
\label{tab:codes}
\footnotesize
\resizebox{0.8\textwidth}{!}{%
\begin{tabular}{l|l|l|r|r}
\hline
\rowcolor{gray!35}\multicolumn{2}{c}{ } & & \multicolumn{1}{c}{\textbf{\# Interviews}} & \multicolumn{1}{c}{\textbf{\# Survey }} \\
\rowcolor{gray!35}&\textbf{Region} & \textbf{Participants' IDs (Interviews)}& \multicolumn{1}{c}{\textbf{(total: 27)}} & \multicolumn{1}{c}{\textbf{(total: 193)}} \\\hline\hline

\rowcolor{gray!40}\multicolumn{1}{r}{~}& \multicolumn{2}{r|}{Participants who mentioned at least one Region in Inter-personal} &
  26 &
  162 (84\%) \\ \hline

\multirow{4}{*}{Quad1}
& Cooperation            &  P2, P3, P4, P5, P7, P8, P13, P15, P17, P20, P21 & 11 & 15 (8\%)\\\cline{2-5}
& Perceived Contribution & P5, P6, P9, P11, P17, P18, P21, P22, P25         & 9  & 57 (30\%)\\ \cline{2-5}
& Recognition            & P1, P9, P13, P22, P23, P25, P27                       & 7  & 29 (12\%)\\  \cline{2-5}
\rowcolor{gray!10}\multicolumn{1}{r}{~}& \multicolumn{2}{r|}{Participants who mentioned at least one Region in Q1}& \textbf{19} & \textbf{93 (48\%)}\\\hline\hline

\multirow{4}{*}{Quad2}
& Advancement           & P1, P10, P12, P16, P20, P21, P22, P23, P24 & 9 & 8 (4\%)\\\cline{2-5}
& Performance           & P4                                         & 1 & 0 (0\%) \\\cline{2-5}
& Factual Contribution  & P2, P6, P12, P14, P18, P19, P20, P21       & 8 & 74 (38\%)\\\cline{2-5}
\rowcolor{gray!10}\multicolumn{1}{r}{~}& \multicolumn{2}{r|}{Participants who mentioned at least one Region in Q2}& \textbf{15} & \textbf{81} (42\%)\\\hline\hline

\rowcolor{gray!40}\multicolumn{1}{r}{~}& \multicolumn{2}{r|}{Participants who mentioned at least one Region in Intra-personal} &
  11 &
  49 (25\%) \\ \hline\hline

\multirow{3}{*}{Quad3}
& Self-Development  & P7, P16, P18, P19, P20, P21, P24 & 7  & 19 (8\%)\\\cline{2-5}
& Creativity        & P10                              & 1  & 2 (1\%) \\\cline{2-5}
\rowcolor{gray!10}\multicolumn{1}{r}{~}& \multicolumn{2}{r|}{Participants who mentioned at least one Region in Q3}& \textbf{8} & \textbf{21} (11\%)\\\hline\hline

\multirow{3}{*}{Quad4} 
& Satisfaction  & P1, P5, P10, P16, P21, P26 & 6  & 14 (7\%)\\ \cline{2-5}
& Security      & P19, P24              & 2  & 17 (7\%)\\ \cline{2-5}
\rowcolor{gray!10}\multicolumn{1}{r}{~}& \multicolumn{2}{r|}{Participants who mentioned at least one Region in Q4}&  \textbf{7} & \textbf{30 (16\%)} \\ \hline\hline

%\rowcolor{gray!20}\multicolumn{1}{r}{~}& \multicolumn{2}{r|}{Participants who mentioned at least one Region from Affect} &  25 & 49 (25\%) \\ \hline

%\rowcolor{gray!20}\multicolumn{1}{r}{~}& \multicolumn{2}{r|}{Participants who mentioned at least one Region from Achievement} &  17 & 95 (49\%) \\ \hline

\hline

\end{tabular}}

\resizebox{0.75\textwidth}{!}{%
\tiny
\begin{tabular}{r}
\textbf{Quad1}: Interpersonal x Affect; \textbf{Quad2}: Interpersonal x Achievement; \textbf{Quad3}: Intrapersonal x Achievement; \textbf{Quad4}: Intrapersonal x Affect\\
The total per quadrant is not the sum of the regions since the participants often provided an answer that was categorized into \textbf{more than one region}.
\end{tabular}
}
\vspace{-0.2cm}

\end{table*}

% ---------------- Quad1 ---------------------
\subsubsection{Quad1: Interpersonal $\times$ Affect}

%\boldification{**Dries said Interpersonal–affect (Quad1) means feelings and perceptions that characterize a career in the world external to the career actor’s ‘‘self''. Quad1 includes these three regions. We had 19(I)+94(S) in this quad. TableX shows this breakdowns**}

The first quadrant in Dries et al.'s model~\cite{dries2008career} is defined by two dimensions: (1) Interpersonal, which represents an individual's relationships with the outside world; and (2) Affect, which represents internal feelings and perceptions that characterize success. This quadrant contains three distinct regions of meaning: Cooperation, Perceived Contribution, and Recognition. 
%Success meanings associated to this quadrant were mentioned by 19 (70.4\%) interviewees and 93 (48.2\%) survey respondents, as described in Table \ref{tab:codes}.

%\boldification{**Have contacts in several different communities, Help the community continuity and evolving and Be able to increase the community's inclusiveness}

\textsc{Cooperation} (Figure~\ref{fig:carrer_success}.I(a)) is defined as working with others (peers, superiors, subordinates, clients, etc.). The collaborative nature of OSS relates to this region as OSS contributors work together, support their community, and help their peers. In our analysis, we identified five categories, which we explain next. 

Success included building social capital, i.e., \textit{``having contacts in several communities''} as it allows identifying sources of help quickly when necessary (P8, P17, P21). It also includes being able to contribute to \textit{``community sustainability"}, so it can be ``as great as it can possibly be" (P3) and ``more diverse and more inclusive'' (P13). ``Bringing people together'' (P8) to \textit{increase the community's inclusivity} was also repeatedly mentioned as a factor of success. Participants often mentioned individual success as part of the community's success: ``having a healthy community is probably the most important thing" (P4) and ``the sign of a healthy open source project is where everybody feels like their voice is heard and their opinion matters" (P7).

%, as illustrated by P7 and P8: ``to help shine a light on the things that other people are ignoring" (P8) and ``the sign of a healthy open source project is where everybody feels like their voice is heard and their opinion matters" (P7).
%. As mentioned in the interviews ``having a healthy community is probably the most important thing" (P4) and 

%\boldification{**Provide opportunities for contributors to grow}

The cooperation aspect of OSS was also highlighted when participants defined their success as the ability to support others' success by \textit{``providing opportunities for contributors to grow''} (P7) and ``become more present and productive" (P15) by ``giving everybody the opportunity [to climb] the contributor ladder'' (P7).
%, because \textit{"an unhealthy symptom would be the same person in literally every single place"} (P4). 

%\boldification{**Mentor and develop the team}

Participants also cited success as being a \textit{mentor} who is ``friendly, didactic, and receptive to increase contributions" (P2 and P20), ``who [neither] burn themselves out, [nor act as] the hero in the situation" (P15). An OSS mentor plays a crucial role in collaborative communities and influences the degree to which a newcomer relates to an OSS community and identifies with it~\cite{carillo2017makes}. Indeed, our participants mentioned that newcomers need to ``feel they are heard" (P3), and that successful mentors \textit{develop the team} by ``let[ting] people participate" (P4) and ``being open to new ideas, whether that could be coding, helping to figure out what the roadmap is, identifying features, identifying bugs, kind of all those things coming together" (P4).

%\boldification{**Dries said Perceived Contribution is serving society through work. in OSS this translates to outreach, and detailed in this 3 things (Fig2-Quad1-Reg1.b)...as explained by these quotes}

\textsc{Perceived Contribution} (Figure-\ref{fig:carrer_success}.I(b)), according to Dries et al.~\cite{dries2008career}, equates with serving society. In the context of OSS, our participants mentioned perceived contribution from the perspective of \textit{outreach}---i.e., ``impact on people in the world" (P11). Participants considered themselves as successful when the product they contribute to has ``high adoption"(P9), ``produce[s] value for the people" (P17), %``leave[s] a legacy"(P25), 
and makes people's lives easier" (P5). 

%and \textit{"gets massive usership"} (P6), also being \textit{"reused by other developers"} (P21). %The outreach was mentioned also as \textit{"leaving it [the contribution] as a legacy"} (P25), \textit{"something that makes people's lives easier"} (P5) and \textit{"produce value for the people"} (P17). 

\textsc{Recognition} (Figure-\ref{fig:carrer_success}.I(c))---or being adequately rewarded and appreciated for one’s efforts or talents~\cite{dries2008career}---was also mentioned by our participants. P13, for example, defined success as \textit{``being recognized by the community and the project's stakeholders."} P1 considered recognition as awareness that ``the maintainer of these projects know that they can come to [participate] as a subject matter expert" (P1).

% and people \textit{"knowing that the key stakeholders in your thing know you do the thing"} (P1).

% ---------------- Quad2 --------------------

\subsubsection{Quad2: Interpersonal $\times$ Achievement}

%\boldification{**Interpersonal–achievement (Quad2) means real accomplishments in the world external to the actor’s ``self'' and includes these three regions. We had 16(I)+80(S) in this quad. TableX shows this breakdowns**}

As per Dries et al.~\cite{dries2008career}, this quadrant includes accomplishments external to the actor’s self across three regions: Advancement, Performance, and Factual Contribution. 
%Success meanings associated to this quadrant were mentioned by 15 (55.6\%) interviewees and 81 (42.0\%) survey respondents, as we show in Table \ref{tab:codes}.

%\boldification{**Dries said advancement is progress and growth. our data shows this translates to these 4 things (Fig2-Quad2-Reg2.a)...as explained by these quotes}

%We classified as \textsc{Recognition} the cases in which success was mentioned as having their name well-known, and as \textsc{Advancement} when success referred to something related to working at a recognized company or a well-known project. 

\textsc{Advancement} (Figure-\ref{fig:carrer_success}.II(a)) is defined as progressing and growing in terms of level and experience. In the OSS context, this relates to \textit{influencing decisions} about the product, being [part of] an \textit{``influential community that is well recognized, a community that you say the name and people know what is}'' (P21), \textit{receiving job offers}, ``writing [one's] own ticket" in one's career (P12), receiving a \textit{salary increase}, or \textit{achieving a top-level position}. ``Money'' in some cases represented growth (e.g., ``salary going up'' (P16)), which differs from some other cases in which money represented a way of earning a living from OSS, which we classify as \textsc{Security}. 

%Advancement was also mentioned as "being able to influence decisions about the product road map" (P7, P23). 

% In our analysis, we found that success meanings in this region are related to one's own progress and growth.

%\boldification{**Dries said performance is achieving verifiable results. in OSS this translated collectively achieving a result(release schedule) (Fig2-Quad2-Reg2.b)...as explained by these quotes}

The \textsc{Performance} (Figure~\ref{fig:carrer_success}.I(b)) region is defined as attaining verifiable results and meeting set goals~\cite{dries2008career}. In our context, this translated to having a \textit{plan for project releases} ``depending on what the goals of the project are, such as working on a new release every six months" (P4). Project planning activities demonstrated the relation of the actor to the external world (interpersonal dimension), as explained by P4: ``if [one is] not making [the release], [they are] letting a lot of people down".

%\boldification{**Dries said factual contribution is individually contributing something tangible to the collective. in OSS this is translated to outreach (produce value, cause impact) (Fig2-Quad2-Reg2.c)...as explained by these quotes}

\textsc{Factual contribution} (Figure-\ref{fig:carrer_success}.II(c)) is about individual contributions to the collective~\cite{dries2008career}. An indication of success in this region includes \textit{bringing contributions to an OSS project}, by ``getting a change that you wrote accepted'' (P12), including ``a code change, a documentation change... [or otherwise] getting something you made merged'' (P12). Besides code contributions, interviewees mentioned implementing ideas or any type of revisions or contributions to the project, as well as ``actively reviewing and looking at what people are suggesting" (P2). Contributions can also represent something tangible, such as achieving financial gains when ``selling the platform'' (P6) or when having a ``ventured organization'' (P6). 

% ---------------- Quad3 --------------------

\subsubsection{Quad3: Intrapersonal $\times$ Achievement}

%\boldification{**Intrapersonal x Achievement (Quad3) means Real accomplishments of the career actor’s ‘‘self'', and includes these two regions. We had 9(I)+21(S) in this quad. TableX shows this breakdowns**}

Dries et al.~\cite{dries2008career} describe this quadrant as including real accomplishments of the actor’s ``self''. It contains two distinct regions of meaning: Self-Development and Creativity. 
%Success meanings associated to this quadrant were mentioned by 8 (29.6\%) interviewees and 21 (10.9\%) survey respondents, as we show in Table \ref{tab:codes}.

%We found success meanings from the Self-Development region, as learning new skills, being a maintainer or able to make changes in any size and any type of OSS project, and receiving enough OSS job offers to be able to choose next step in the OSS career; and from the Creativity region, as creating new knowledge and implementing ideas. 

%\boldification{**Dries said self-development is reaching full ability through self-management of challenges and learning experiences. in OSS this is translated to those 4 things) (Fig2-Quad3-Reg3.a)...as explained by these quotes}

\textsc{Self-Development} (Figure-\ref{fig:carrer_success}.III(a)) is defined as realizing one's potential through self-management of challenges and learning experiences~\cite{dries2008career}. This has been a classic motivation for contributing to OSS~\cite{hertel2003motivation, harsworking, lakhani2003hackers}. However, success definitions mentioned by the interviewees go beyond \textit{``learning new skills''} (P16). They also include the path to receive a promotion, as stated by P20: ``I reviewed other people's code to improve my review skills to become a maintainer,'' and \textit{be prepared to serve as a key project member} by ``being a mature reviewer and contributor" (P2) \textit{``capable of effecting change in an open source project, from the small to the large''} (P7). 

%It is worth to mention that we classified as Factual Contribution when the person mentioned bringing contribution to an OSS project, and we classified as Self-Development when the person mentioned becoming capable to do so.%, "make contributions, put that on his resume as a reference and start having job opportunities, that is success (P19) 

% \boldification{**Dries said creativity is creating something innovative. in OSS this is translated to create new knowledge and implement ideas) (Fig2-Quad3-Reg3.b)...as explained by these quotes}

\textsc{Creativity} (Figure-\ref{fig:carrer_success}.III(b)) is about making something innovative and extraordinary~\cite{dries2008career}. We found this to mean the freedom to \textit{``create new knowledge''} (P3), but also as \textit{``propagating ideas''} (P3). Creativity is relevant to the OSS context as individuals from innovative communities have greater opportunities to express themselves and experience a sense of accomplishment~\cite{lakhani2003hackers}.

% ---------------- Quad4 --------------------
\subsubsection{Quad4: Intrapersonal $\times$ Affect}

%\boldification{**Intrapersonal x Affect (Quad4) means Feelings and perceptions that characterize a career of actor’s ‘‘self'', and includes these two regions. We had 8(I)+30(S) in this quad. TableX shows this breakdowns**}

The Intrapersonal $\times$ Affect quadrant includes feelings and perceptions that characterize the career of an actor’s ‘‘self'' ~\cite{dries2008career}, which contains two regions: Satisfaction and Security. 
%Success meanings associated to this quadrant were mentioned by 7 (25.9\%) interviewees and 30 (15.5\%) survey respondents, as we show in Table \ref{tab:codes}.

%We found success meanings from the Satisfaction region, as being happy, having more friends, participating on the OSS scene and being able to express and having your ideas heard; and from the Security region, as becoming a paid contributor and having money to sustain yourself from OSS. 

%\boldification{**Dries said satisfaction is achieving happiness and personal satisfaction. in OSS this is translated to those 4 things (Fig2-Quad4-Reg4.a)...as explained by these 4 things. The money thing here is different from salary increase, which means Advancement}

\textsc{Satisfaction} (Figure-\ref{fig:carrer_success}.IV(a)) is about achieving happiness and personal satisfaction, either in the family or in the work domain~\cite{dries2008career}. Participants mentioned satisfaction as \textit{``being happy"} (P1, P16, P26), which also included \textit{``being able to express yourself"} (P10). They talked about their \textit{sense of belonging} and ``need for emotional inclusion" (P16), the importance of ``participating in the world that is being created" (P10), and having \textit{``a ton of friends} and people [who they] would hang out with or chat with, about non technical stuff" (P5). 

%\boldification{**Dries said security is related to financial or employment needs. in OSS this is translated to live from OSS) (Fig2-Quad4-Reg4.a)...as explained by these quotes. The money thing here is different from salary increase, which meant Advancement}

\textsc{Security} (Figure-\ref{fig:carrer_success}.IV(b)) means meeting one’s financial or employment needs~\cite{dries2008career}. Participants characterized success as the ability to \textit{make a living from OSS}--- to ``receive money as an OSS developer" (P24) and ``prioritize what [financially] sustains you''~(P19).

%We noticed that "money" concept reappeared here, from a different perspective and a different meaning: 

%\subsubsection{Considering yourself a successful contributor}

\MyBox{Success is a multifaceted and complex concept, including both objective metrics and subjective perceptions of accomplishments. 

%Despite its vastly different philosophy of work, we could classify the perceptions of success reported by the OSS contributors from our study into the same regions that emerged from Dries et al.'s \cite{dries2008career} study
%Despite its vastly different philosophy of contribution, OSS can provide similar avenues for success as described by Dries et al.'s model.%in more ``traditional models".
} 

%\MyBox{\textbf{What is success in OSS?} Success in OSS is a multifaceted and complex concept, which involves both feelings and objective accomplishments and have different meanings for each person. The concept of success grounded in our data could be linked to all the regions of a more general model. This evidences that, although OSS ecosystem is different from more `traditional' ones, the meaning of success is similar. %Lastly, we also found that success for a single person have meanings crosscutting the dimensions of the model, showing how complex it is to measure it.

%We also found that success is a dynamic concept \wnote{why? because?}(e.g. money can represent one's Security as earning own sustain or Advancement as salary increase).

%The components of success may include receiving enough OSS offers to make possible choose the job, earn your sustain from OSS, be happy, learn and create new knowledge. Bring contributions to an OSS project, being recognized for any contribution you do, working for a recognized project or a well-known company, achieving a high level in hierarchy, becoming a key member, increasing the salary and influencing decisions about the product road map. Having personal outreach and contacts in several different communities, but also produce products impact the most possible amount of users. Be a mentor and provide opportunities for members to grow, help the community to be more inclusive, continue existing and evolving.

\vspace{-0.0cm}
\subsection{Survey analysis}
\label{sec:results_segmented}

%\boldification{***goal of survey was to triangulate our def of success. We didn't find any new definition. Here are the results in broad strokes***}

As explained in Section~\ref{sec:phase2}, we conducted a survey to triangulate the definitions of success we identified from the interviews, expanding our population and exploring whether we could find any new definitions of success. We qualitatively analyzed the 193 answers to our survey open question. Similar to interviews, the participants often provided multiple definitions, which could be categorized into more than one region from Dries' model \cite{dries2008career}. However, no new category emerged from the survey analysis.

%\review{We used the codebook created after the interviews as thee seed to the analysis of the survey questions. We diligently analyzed the answers to classify the answers and to identify any new perceptions of success that did not emerge from the interviews. At the end of the process we could not find anything that could not be classified using the existing codeset.} Thus, the 26 codes from our \textit{interviews were sufficient to encompass the definition of success from the survey respondents}.
%Although we had the same level of attention to define the perceptions of success that would not be previously classified by the interviews, t}he survey analysis did not reveal any new categories. 

In this section, we look deeper into the survey results to understand the prevailing definitions of success among our respondents and across different demographics. When presenting the results, we use supplementary and corroborative counting of the survey responses to triangulate the qualitative analysis of the definitions of success~\cite{hannah2011counting}.

%\textbf{The Interpersonal dimension} 
\textbf{The dimensions of success.} 
The majority of respondents defined success in terms of a relationship with the external world (Interpersonal) rather than the actor's self (Intrapersonal), accounting for 84\% vs. 25\% of respondents. For the Interpersonal dimension, respondents identified success across both ends of the Affect and Achievement spectrum---25\% were related to the Affect dimension and 49\% were related to Achievement. When considering definitions related to the Intrapersonal dimension, none of the regions were mentioned by more than 10\% of the respondents. This preponderance of definitions related to the Interpersonal side could be due to the collaborative nature of peer-production sites such as OSS, where contributing to a common good and being recognized for it have been cited as key motivation factors~\cite{von2012carrots,gerosa2021motivation,hertel2003motivation,harsworking,roberts2006understanding}.

In fact, \textsc{Factual} (38\%) and \textsc{Perceived Contribution} (30\%) were the most mentioned regions, followed by \textsc{Recognition} (12\%). None of the other regions across all quadrants had more than 10\% of responses. These responses reflect that, in OSS, while contributions matter, the way that others (community, peers, society) value the contributions is also an important indicator of success. 

Respondents who identified \textsc{Factual Contribution} as a definition of success emphasized that the number, size, and frequency of contributions can be objective concepts to quantify a significant contribution to the community. They defined success as \textit{``finding a way to sustainably contribute"} (S25), or being \textit{``someone who is able to regularly contribute"} (S11) and \textit{``spending time on the project often"} (S68). A successful contributor is one who provides \textit{``a wide spectrum of contributions"} (S6). Moreover, respondents identified various types of contributions for contributors in different project-centric or community-centric roles \cite{trinkenreich2020pathways}, as mentioned by S2: \textit{``Successful contributors add or change major features, and organize the community"}.

Those who considered \textsc{Perceived Contribution} as success emphasized the importance of their contribution, such as publishing and maintaining software that is used by and useful to a lot of people. According to S136, the perceived value of their contribution could be measured by \textit{``how many people have used the OSS code and how much value has it created"}. Some of these definitions of success in OSS included: \textit{``someone who publishes and maintains software that is useful for a lot of people or for the user community"} (S3) and \textit{``when the software solves and helps real world problems"} (S169).

Finally, our respondents reflected many different perceptions of success related to \textsc{Recognition} in their community; which included \textit{``having a high number of stars on the own repository in GitHub''} (S58 and S109), \textit{``receiving donations''} (S21), and \textit{''being invited for conference invites/talks''} (S16).

\begin{figure*}[htb]
     \centering
     \includegraphics[width=0.7\textwidth]{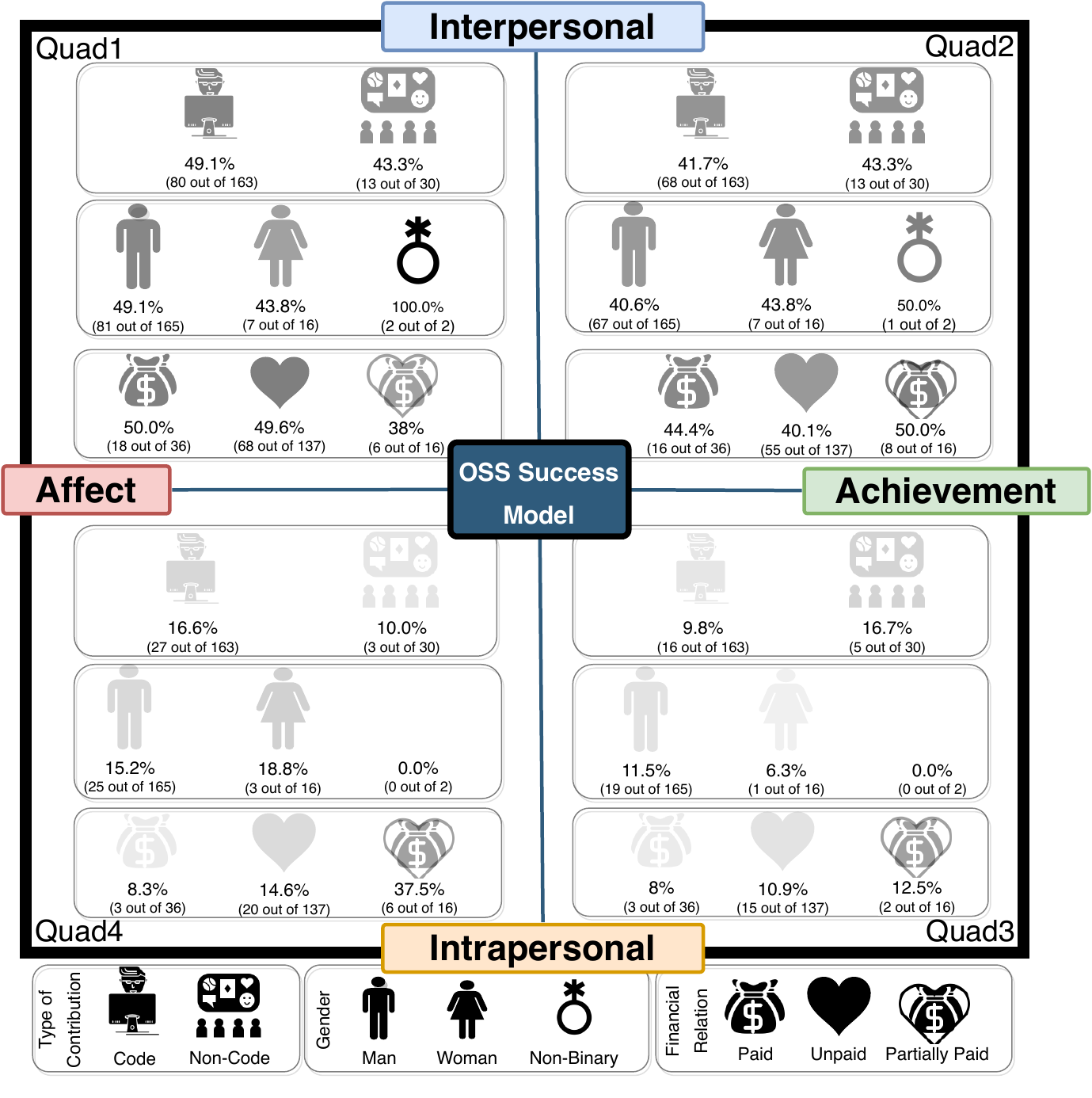}
     \vspace{-0.4cm}
     \caption{Subgroup analysis of the meanings of success. The opacity of the icons represents the percentage of each group for the quadrant. Darker means a higher and lighter a lower percentage. Some respondents provided answers about success that accounted for more than one quadrant.}
   %  \vspace{-0.5cm}
     \label{fig:segmented.analysis}
 \end{figure*}

\boldification{*** demographics and success definition ***}
\textbf{Demographics and the meaning of success} As recent literature has shown, the OSS community is becoming more diverse in terms of the gender of contributors, types of contributions, and financial rewards~\cite{carillo2017makes,trinkenreich2020pathways}. We took a deeper look into these demographic subgroups with respect to their definitions of success. Understanding how different demographics perceive success can help us create mechanisms to better support diverse contributors and improve the state of diversity in OSS. Figure \ref{fig:segmented.analysis} illustrates the definition of success for each demographic subgroup. 
%The figure shows the analysis of the stratified data according to type of contribution, gender, and financial relationship with the project. 
The percentages in the figure reflect the number of participants who mentioned any meaning under each quadrant per subgroup. For example, 80 participants who identified themselves as code contributions reported at least one meaning of success categorized in Quad1. Therefore, given there were 163 code contributors, 49.1\% of the code contributors in our sample associated success with Cooperation, Perceived Contribution, or Recognition (Quad1).

%\textit{Gender.} 
From the 193 survey respondents, 165 identified as men, 16 as women, and 2 as non-binary. The gender distribution of our respondents matches that of those reported in other OSS studies (\cite{vasilescu2015gender,robles2016women,singh2019women}). We dropped from this analysis the 10 respondents who did not disclose their gender. Comparing the distributions of definitions reported by men and women, we could not find statistically significant differences between the two groups either in terms of quadrants or regions. As illustrated in Figure~\ref{fig:segmented.analysis}, both men and women more frequently mentioned success definitions classified in the Interpersonal quadrants (Quad1/Quad2) than those in Intrapersonal (Quad3/Quad4).

%\textbf{Type of Contributions and Monetary Rewards} 
Our survey included answers from 163 \textit{coders} and 30 \textit{non-coders}, i.e., those who work only on non-code related activities (e.g., advocacy, license management, technical writing). We could not find statistically significant differences between the distribution of answers from the two subgroups. We could also not find statistically significant differences when sub-grouping based on compensation (paid vs. unpaid). The statistical test results including the p-values of these comparisons are in the supplementary material.

\MyBox{The Interpersonal dimension plays a dominant role in the definition of success, in which factual and perceived contributions are the most referenced, followed by recognition. Contributors across different demographic groups---gender, contribution type (code vs non-code), and compensation (paid vs. non-paid)---report similar perceptions of success.}
\vspace{-0.3cm}

\section{Discussion}
\label{sec:discussion}

%In this section, we discuss our main findings.

\subsection{Success is multifaceted and hard to measure}

%\boldification{***We found that success is a fluid concept and may be composed of multiple dimensions. People reported different meanings, and for a single person the meaning spam across multiple dimensions***}

Success in OSS is a complex concept with multiple dimensions. Our participants reported different definitions for success, encompassing all the regions of the Dries et al. model \cite{dries2008career}. Even a single person's understanding can span multiple dimensions. Therefore, the dominant view of successful OSS contributors as code ``hackers'' \cite{lakhani2003hackers} is inadequate, even from the point of view of coders. Success in OSS is a nuanced, multifaceted concept that goes well beyond becoming a core member or a maintainer. 

%Our data could be arranged within a general success model, which shows that, although OSS is a different endeavor, the way that people can achieve success is the same, opening opportunities for people with different goals.

%Money, for example, may represent \textsc{Security} when related to a way to live from OSS, and \textsc{Advancement}, when related to career growth (salary increase)

%Dries et. al \cite{dries2008career} reported that salary was the construct they had more difficult to fit in the model, as it could represent \textsc{Security}, \textsc{Advancement} (as we also found), and \textsc{Recognition}, when their participants mentioned to receive a ``fair salary". Although our data did not include ``money" as Recognition, we had another dynamic example for this construct. When contributing for a recognized community, it represents \textsc{Advancement}. 
%When recognition was mentioned as being their own names well-known or having notoriety, it represents \textsc{Recognition}. 

%\boldification{Even concepts that look simple, may have, in fact, different meanings. Every single thing depends on the perspective (subjective vs. objective)}

%\boldification{***Every single thing depends on the perspective (subjective vs. objective) making it hard to measure. ***}

This variety of perspectives makes it challenging to measure success. Even common terms, such as ``contribution,'' can be understood differently. While some people consider a high number of contributions or the frequency of contributions as a measure of success, others relate success to the impact of their contribution---or how it is perceived by the users or the society. Current literature, tools, and project infrastructures unfortunately tend to focus on measuring code-centric contributions (e.g., \cite{roberts2006understanding,huang2016cpdscorer,wang2020unveiling}). However, there are subjective perspectives of success closer to the \textit{Affect} dimension that also need to be measured. For example, how does one measure contributions for those who mentor or work on community building (Quad1)? The benefits of these types of contributions are intangible and by their nature difficult to measure.
In fact, our results show even tangible products, such as money, can represent multiple meanings of success: for instance, \textsc{Security} when related to making a living from OSS, and \textsc{Advancement}, when related to growth (salary increase). Therefore, it is important that researchers and practitioners take a more nuanced approach in developing ways to evaluate success, considering the multitude of profiles and activities that are part of OSS. There is no ``one size fits all" measure of success.
%Even contributions go beyond the objective way of measuring, when including the subjective side of how the contribution is perceived.

%It is imperative that we  move beyond the current code-centric approaches to recognize and reward OSS contributors. 

%This model can be used by the communities as well as researchers to better support the goals of different success-seekers in OSS. 
%When knowing how their contributors perceive success, the communities can cultivate those success attributes within their members~\cite{li2020distinguishes}. This might help make OSS more diverse by attracting a diverse set of people---both in demographics as well as skills. Novice professionals can see different ``pots of gold'' they can have through working in OSS and be attracted to OSS projects, increasing contributions and diversity.

%The complexity of the success concept makes the phenomenon challenging enough to arouse the interest of researchers. %OSS practitioners may also reflect and think about extending the existing approaches based on badges and counting to include other ways to recognize successful contributors.

%present ways of measuring the project or product success (e.g. \cite{crowston2003defining, subramaniam2009determinants, piggot2013healthy}). Studies have been focused on measuring the meanings related to the Achievement end, mostly code-centric contributions (e.g. \cite{roberts2006understanding, huang2016cpdscorer}, and their performance on a set of programming tasks~\cite{bergersen2014construction}. 

\subsection{Coders \& non-coders look for the same pot of gold}

While coders and non-coders may contribute differently to OSS, they perceive success in similar ways. Our analyses (Figure~\ref{fig:segmented.analysis}) show that both coders and non-coders often mentioned definitions that relate to the \textit{Interpersonal} dimension. 
%More specifically, \textsc{Factual Contribution}, \textsc{Perceived Contribution}, and \textsc{Recognition} were highly cited ( Table~\ref{tab:codes_survey}). 

Coders and non-coders perform different roles and have different career pathways in OSS~\cite{carillo2017makes, trinkenreich2020pathways}. These pathways may include not only code-centric, but also community-centric activities, including advocacy, community building, mentorship, and technical writing. These activities are important for projects' sustainability and growth, but are currently not well-recognized.

Therefore, showing that there are multiple ways to achieve success is important, regardless of their roles. To do so, current strategies and metrics to support contributors need to be adapted to consider the multitude of success definitions to include activities not directly related to code. For example, coders gain recognition from having their names in a ``credits'' file or badges in their profiles, but non-coders are commonly overlooked because their activities are harder to quantify~\cite{barcomb2018uncovering}. Identifying ways of showing \textsc{Recognition} for non-coders is important future work. 

%Still, even knowing that the main ``pot of gold'' for coders is the same, it is important to consider the diverse meanings of success. Recognizing the different natures of contribution, and being careful when using myopic rankings and badges to only certain types of contributors. 

%follow different bands of the rainbow, the end pot of gold remains the same.

\subsection{Subjective definitions of success is prevalent}

In our study, both men and women mentioned success definitions related to the \textit{Affect} and \textit{Achievement} dimensions, and at similar rates. Contrary to research in other domains~\cite{dyke2006we, cho2017south, porter2019physics} that found that men relate success to tangible and objective outcomes, the men in our study often provided subjective meanings of success.
%Contrary to the research in other domains~\cite{dyke2006we, cho2017south, porter2019physics}, meanings of success related to \textsc{Cooperation} and \textsc{Perceived Contribution} (Table~\ref{tab:codes_survey})---which are part of the Affect end---were more cited by men than women. The literature shows that men relates success to tangible and objective outcomes. 

We hypothesize that the nature of OSS defines how people see success in this context. OSS is an open collaboration community~\cite{forte2013defining}, in which collective work is central to the success of projects. Additionally, altruism, reciprocity (giving back), and maintaining high-quality social bonds are common motivations to contribute to OSS~\cite{von2012carrots,gerosa2021motivation}. These motivations relate to working together to create better software and for the greater good. Individuals who are attracted to an open-collaboration community may attribute a high value to these dimensions.

%\draft{\subsection{}

%

%The reasoning for that can be OSS is all about cooperating with others, and the community valuing the contributions that we do.

%\subsection{Money matters (if you earn some)}
%\textcolor{blue}{AS: boring...yep better to del since its 1/3 of 16 people very low number of people}
%Partially paid contributors cited more living from OSS as a meaning of success than the paid and unpaid contributors. This can mean that partially paid contributors are interested on moving to a full-time job in OSS and make OSS be their main source of money. While that, paid and unpaid contributors cited more success meanings related to \textsc{Factual Contribution}, \textsc{Perceived Contribution}, and \textsc{Recognition}. 

\section{Implications of Results}
\label{sec:recommendations}

There are several ways our results can inspire communities and researchers to engage OSS contributors.
\vspace{0.2cm}

\subsection{Recommendation for Communities}

% \boldification{**Implication 1: Communities can support their members to achieve success now what means being successful for them. When feeling advance toward their goals, contributors can stay more, reducing the high rotation that happens in OSS.}

%The perceptions of success reported here can help OSS communities and organizations better support contributors on achieving the success for which they aim. 

% \boldification{**Implication 2: Novice professionals, when aware of the many different ways of being successful in OSS, can be more attracted to join OSS projects.}

Open source offers different ``pots of gold" for different types of contributors. OSS contributors are diverse, have different motivations to join OSS and have different definitions of success. Our results highlight these differences. Being aware of the diverse success definitions can allow a diverse set of developers to be inspired to join OSS and find others who value similar aspects of success. 

Our results can make OSS communities aware that individuals have diverse backgrounds and perceptions of success and may need different engagement strategies. By recognizing that success is multifaceted, communities can leverage our (sub-)categories to support the growth of individuals who hold different success definitions. By understanding what contributors they seek, leaders can cultivate practices that highlight the success attributes in their projects to improve retention and turn over rates~\cite{kimmelmann2013career}. 

%%%Quad1 - Cooperation, Perc Contribution and Recognition
Communities, for example, can foster engagement of contributors who define success in terms of increasing their personal networks and \textbf{\textsc{Cooperation}}~\cite{borges2019developers,ingram2020software}. They can do so by organizing meetups to increase contributors' social capital (Fig.~\ref{fig:carrer_success} I(a)). Communities can organize Hackathons~\cite{trainer2016hackathon} or participate in ``Summer of Code" programs~\cite{SILVA2020110487,silva2017students, silva2020theory}, which offer the contributors opportunities to help with ``outreach'' (\textbf{\textsc{perceived Contribution}}) (Fig.~\ref{fig:carrer_success} I(b)) and improve the sustainability of the project (Fig.~\ref{fig:carrer_success} I(a)) by mentoring and onboarding new members. 

Communities can employ different \textbf{\textsc{recognition}} programs to value different types of contributions~\cite{trinkenreich2020pathways} and engage those who perceive success as ``being recognized''(Fig.\ref{fig:carrer_success} I(c)). For example, in addition to traditional metrics such as code commits, communities can award  contributors who participate by answering questions and discussing issues~\cite{ducheneaut2005socialization}. 
%%% Quad2 - Factual Contribution, Performance, Advancement
Recognizing different types of contributions can engage those who value \textbf{\textsc{factual contributions}} (Fig.~\ref{fig:carrer_success} II (c)).

\textbf{\textsc{Performance}} and merit-based badges~\cite{wu2018network} can  be used to recognize contributions and community building~\cite{copenhaver2017digital,papoutsoglou2020modeling} (Fig.~\ref{fig:carrer_success} II (b)). Communities should make explicit their criteria and rules for promotion~\cite{picot2015social}, making contributors aware of what is expected in terms of skills and contributions to achieve their \textbf{\textsc{Advancement}} goals (Fig.~\ref{fig:carrer_success} II (a)).
%%% Quad3 - Self-Development and Creativity

Communities are encouraged to prepare manuals and iterative learning modules and provide skill-specific mentoring~\cite{fagerholm2014role} to help with continuous learning and \textbf{\textsc{Self-Development}} (Fig.~\ref{fig:carrer_success} III (a))~\cite{fiesler2017growing}. The training content should not only cover technical topics, but also how to improve other skills (e.g., communication, networking). Contributors who value \textbf{\textsc{Creativity}} can be engaged via badges that highlight different skills  (Fig.~\ref{fig:carrer_success} III (b)) \cite{copenhaver2017digital} or by building and sharing new knowledge for online training of new developers \cite{steinmacher2018let}. 

%%% Quad4 - Satisfaction and Security

As sense of belonging is directly related to job \textbf{\textsc{Satisfaction}} (Fig.~\ref{fig:carrer_success} IV (a))~\cite{lim2008job}, communities can promote inclusivity events~\cite{izquierdo2018openstack,canedo2020work}, ultimately aiming to reduce contributors' loneliness and alienation, and providing social support for mental health. Communities can help contributors avoid burnout, which can negatively affect satisfaction, well-being, and happiness~\cite{graziotin2018happens,nowogrodzki2019support}, by further supporting key members~\cite{raman2020stress}. Finally, to support and retain contributors who wish to achieve financial \textbf{\textsc{security}} from OSS, communities can explicitly state their partnerships with companies, offer ``bounties" as payment per issue solved~\cite{zhou2021studying}, or join onboarding programs (such as Google Summer of Code) that compensate participants \cite{SILVA2020110487,silva2017students,silva2020theory}.

%to offer developing (all or part of) the software they need as they fund the contributors (as Apache, Linux, SendMail, Mozilla and many others do) \cite{roberts2006understanding}

\vspace{0.3cm}

\subsection{Implications for Researchers}

%\boldification{**Implications 3: Researchers can explore how to get successful in OSS, ways to measure and help contributors.}

The multitude and nuance of definitions of success can serve as input for different research directions. It is important to find ways to support the growth of people whose background is not related to software development. Their activities are harder to quantify, given that they usually do not leave traces on project repositories. This may pose challenges beyond proposing metrics and toward proposing changes in terms of how these activities are performed, logged, and weighted. This may have additional impact on topics like mining and creating virtual resumes for hiring purposes~\cite{sarma2016hiring}, recommending mentors~\cite{canfora2012going}, and providing paths to becoming central to the project~\cite{zhou2012make,agrawal2018we}.

%, including ways to recognize their contributions and 
% badge, visual rèsumè, metrics, mentor recommendation

%This  is  a  call  of  action  to  our  fellow researchers:join us in helping make OSS diverse by finding different ways to  support  diverse  individuals  with  diverse  backgrounds  and motivations who have diverse definitions of success

%supporting OSS communities on putting plans in practice, defining measurement to control and improve actions, while managing the contributors' expectations about their success perceptions.

\vspace{0.5cm}
\section{Related Work}
\label{sec:relatedWork}

%In this section, we introduce the context of success in the context of software engineering and OSS.

Thus far, the literature has discussed how to make projects successful~\cite{subramaniam2009determinants,midha2012factors,sen2012open}. In this study, our goal is to understand success from the contributors' perspective. People's perceptions of themselves impact their behavior and choices to achieve desired goals~\cite{cantor1986motivation,karniol1996motivational}. Motivations to join and to stay also influence how people behave. In the following, we discuss related work focused on motivation in OSS and on skills needed to be a successful software developer. 

\subsection{Motivation to be an OSS contributor} 

Motivation to be an OSS contributor has been extensively studied since the early 2000s \cite{harsworking,ghosh2002free,lakhani2003hackers,hertel2003motivation,hann2004developers,roberts2006understanding,choi2015characteristics, spaeth2015research, bosu2019understanding}. Von Krogh et al. \cite{von2012carrots} surveyed the literature and aggregated the studies about motivation in OSS published until 2009. They identified that the reasons to join OSS can be summarized into 10 motivation categories, grouped as intrinsic, internalized-extrinsic, and extrinsic. %Intrinsic motivation moves the person to act for the fun or challenge entailed rather than in response to external pressures or rewards~\cite{ryan2000self}. In contrast, extrinsic motivations are based on outside incentives when people change their actions due to an external intervention~\cite{frey1997relationship}. Developers can also internalize extrinsic motivators in a way that they are perceived as self-regulating behavior rather than external impositions~\cite{deci1987support,roberts2006understanding}. 
%These internalized extrinsic motivations include reputation, reciprocity, learning, and own-use. 

More recently, Gerosa et al.~\cite{gerosa2021motivation} identified that while career is an extrinsic motivation relevant to many contributors, intrinsic and internalized motivations explained most of the contributors' motivations. %The study revealed that social aspects (e.g., altruism, kinship, and reputation) are now more represented as motivations than previous surveys. 
Wu et al.~\cite{wu2007empirical} investigated the relationship between motivation and retention, and found that altruism, learning, career, and own-use are the main motivators that influence the intention to continue in the project, which was confirmed by Gerosa et al.~\cite{gerosa2021motivation}. % found that the contributors' most common motivation to stay in OSS are fun, altruism, reputation, and kinship. The current literature discusses the set of motivators that drive contributors to join OSS and stay around, however they did not analyze how they perceive success in this context.} 
%While past work has studied what motivates people to contribute to OSS (e.g., \cite{hertel2003motivation,gerosa2021motivation,von2012carrots}),

Success perceptions and motivation to contribute complement each other, but play different roles. When considering the comprehensive study from Von Krogh et al.~\cite{von2012carrots}, although there is an intersection between definitions of success and motivation factors (e.g., money, ideology, reputation), not all motivation factors map to success definitions (e.g.,``Own-use'') and not all the success definitions map to motivation factors (e.g., ``Have a plan for project releases (Performance)''). 
%On the other hand, there is an intersection between definitions of success and motivation factors. For example, money, ideology, and reputation can be found both in the OSS motivation literature \cite{von2012carrots} and in our model of success. 
In our study, we aim to highlight that OSS offers a multitude of success perspectives that, together with the motivation to join and to stay, should be used to understand and support diverse contributors. Our results can be used in future work to investigate how OSS contributors with different motivations perceive success.

\subsection{The skills of successful individuals} 

Past research have been dedicated to providing answers to the question of what attributes and skills make someone successful in their current profession, using the term ``great'' as a proxy for success.
%Learning new skills is one of the perceptions of success part of Self-Development region, and we then reviewed the literature about the skills to be great.
Li et al.~\cite{li2020distinguishes} conducted a study to identify the characteristics that distinguish ``great'' software engineers. The authors found that the top five characteristics are writing good code, regulating behaviors to account for future value and costs, exercising informed decision-making, avoiding making colleagues’ work harder, and constantly learning. Kalliamvakou et al.~\cite{kalliamvakou2017makes} investigated the attributes of a ``great" manager. According to their study, some level of technical skill is necessary, but they are not as relevant as management skills to guide engineers to make decisions, to motivate them, and to mediate their presence in the organization. Dias et al.~\cite{dias2021makes} presented a conceptual framework to explain how  management, social, technical, and personality attributes are connected. They noted that a great maintainer needs both technical excellence and good communication. Through six interviews, Kimmelmann~\cite{kimmelmann2013career} investigated the technical, social, and personal competencies developers need according to their stage in OSS projects, and claims that these competences can support or hinder a successful career by regulating professional behavior. 

Some research considers core developers as ``elite contributors"~\cite{wang2020unveiling} or code heroes~\cite{agrawal2018we}, who receive commit rights based on trust~\cite{sinha2011entering}. Although code heroes are valuable for OSS projects~\cite{agrawal2018we}, being a core developer is not the only way to be successful. According to Zhou and Mockus \cite{zhou2012make}, newcomers become Long Term Contributors if they start with comments and demonstrate a highly community-oriented attitude. While the theoretical converging lens orients most OSS research efforts towards the project-centric and technical side of OSS project development, our study aims to unveil other perspectives of success beyond the traditional ways to measure success of OSS contributors.

\section{Limitations}
\label{sec:limitations}

%In the following, we discuss validity and reliability of our results from the perspective proposed by Merriam~\cite{MerriamBook}.

\textbf{Internal validity}. The characteristics of our sample may have influenced our results. A great part of our interviewees (11 out of 27) were speakers at an OSS conference and half (13 out of 27) of the interviewees identified as women, even though we did not push toward having an equal gender split. This diversity of profiles helped bring a more diverse perspective about the phenomenon. Our survey, which received almost 200 answers, corroborated our results. The distribution of our survey demographics is similar to the larger OSS population as reported elsewhere ~\cite{GitHubOpenSourceSurvey2017, robles2016women, vasilescu2015gender}.

\textbf{Construct validity}. One threat to construct validity in this work relates to the question about success, which explicitly asks how the respondent defines a successful person in OSS. While the question refers to the individual's perspective, respondents could interpret the question differently and answer from the perspective of a typical contributor. This was not a problem for the interviews, since we would have been able to clarify the question if any the interviewees misinterpreted this question (none did). We believe this threat to be minimal in the survey based on our pilot studies. Moreover, individuals' perceptions about typical and prominent participants in the OSS ecosystem are also relevant in creating a broad understanding of success. The theoretical model~\cite{dries2008career} used to categorize the definitions of success also may pose a threat. However, the model was able to capture the nuances of success in OSS, enabling the researchers to map the concepts to the regions proposed by the model.

\textbf{Survival bias}. Our results reflect the opinion of current contributors who joined OSS and made it past the initial contribution barriers~\cite{steinmacher2015social}. Therefore, to promote diversity in OSS, we acknowledge that additional research is necessary to understand success from the perspective of those who do not make it past the initial barriers and those who are currently not attracted to OSS.

\textbf{Recall bias.} Moreover, as our survey question was open ended, our results could be impacted by either salience bias, where respondents focus on definitions that are prominent or emotionally striking and not necessarily all the factors that matter; or by memory bias, where participants answered questions based on what they can first recall and not necessarily what’s most important to them. 

\textbf{Data Consistency}. Consistency refers to ensuring that the results consistently follow from the data and there is no inference that cannot be supported after the data analysis~\cite{MerriamBook}. The same group of researchers performed the qualitative analysis of interviews' transcripts and survey's responses. We had weekly meetings to discuss and adjust codes and categories until reaching agreement. In the meetings, we also checked the consistency of our interpretations, continually discussing our results based on definitions of Dries et al.'s model~\cite{dries2008career}. All analysis was thoroughly grounded in the data collected and exhaustively discussed amongst the whole team. The team includes researchers with extensive experience in qualitative methods.

\textbf{Theoretical saturation}. A potential limitation in qualitative studies regards reaching theoretical saturation. In this study, we interviewed 27 participants with different backgrounds and perceptions about the studied phenomenon. The participants are diverse in terms of gender, number of years involved with OSS, and highest achieved academic degree. We kept inviting participants until we could not find any new concept for five consecutive interviews. Moreover, we collected answers from 193 respondents about what it means to be a successful OSS contributor, and we could not find any new meanings. Therefore, although theoretical saturation cannot be claimed, we believe that we obtained a consistent and comprehensive account of the phenomenon.

\section{Conclusion}
\label{sec:conclusion}
%Nevertheless, OSS has considerably changed in the last 20 years, from being a generation of community of volunteers that eventually interact with the software industry, to an ecosystem in which industry consortia push OSS projects forward having a significant amount of professional and paid contributors ~\cite{robles2019twenty}. 

% However, developers may have different perspectives of success, and the contributors who perform non-coding activities need to be heard. 

%Without the understanding of what success means for current contributors different people, it is not possible to offer equal support for the different contributors. 

In this paper, we studied how OSS contributors define success. OSS has considerably changed over the last 20 years, from a generation of code-oriented volunteers to an ecosystem in which industry consortia push OSS projects forward with a significant amount of professional and paid contributors~\cite{robles2019twenty}. Our results show that OSS contributors have a broader perspective on success than the narrow focus on code-related activities---which is better supported by current tools and practices.
%Our results show that OSS contributors view success from a broad perspective, which is much broader than the narrow perspective of code-related activities that current tools and practices seem to highlight.

Our study of 27 interviews with well-recognized OSS contributors and a follow-up survey of 193 OSS contributors reveals a multi-faceted definition of success. We found 26 categories of definitions through our interviews and framed them through the theoretical lens of an existing success model~\cite{dries2008career}. Our analysis shows that success includes objective and subjective accomplishments. Even tangibles such as ``money" can have different meanings to different people (e.g., a way to advance in career or a way to secure a living).

In conclusion, we hope our work in revealing the nuanced definitions of success that OSS contributors have can help us find out how to support diverse individuals with diverse backgrounds. 
%Just as a rainbow is not complete without all its bands of color, so are the different types of contributions needed to make OSS successful. 
Let us work together to support the different contribution pathways to help individuals reach that elusive pot of gold at the end of the rainbow. 

%We By uncovering a diverse spectrum of success in OSS, we believe that it is possible to open more avenues for diversity in OSS. Showing that it is possible to achieve different goals, more people may change their opinion about building a career in OSS, which may, ultimately, result in more people joining the movement.

\section*{Acknowledgments}
We thank our interviewees and survey participants for their time. This work is partially supported by NSF (1815486, 1815503, 1900903, 1901031) and CNPq (\#313067/2020-1).
\ifCLASSOPTIONcaptionsoff
  \newpage
\fi
\bibliographystyle{IEEEtran}
\bibliography{references}
%\begin{IEEEbiography}{author1}
%\end{IEEEbiography}
\vspace{3cm} % to make all the bios in the same column
\begin{IEEEbiography}[{\includegraphics[width=1in,height=1.25in,clip,keepaspectratio]{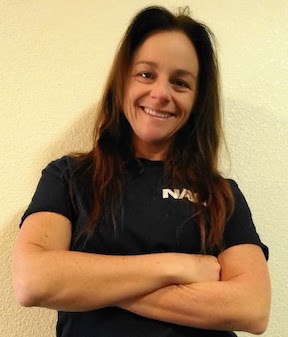}}]{Bianca Trinkenreich} is a PhD student at the Northern Arizona University. She holds a Master of Science in Computer Science from the Federal University of the State of Rio de Janeiro (UNIRIO) and researches about Software Engineering, CSCW, Software and IT Service Quality. Recent projects include the career pathways and motivations of Open Source contributors. \end{IEEEbiography}
\vspace{-1.2cm}

\begin{IEEEbiography}[{\includegraphics[width=1in,height=1.25in,clip,keepaspectratio]{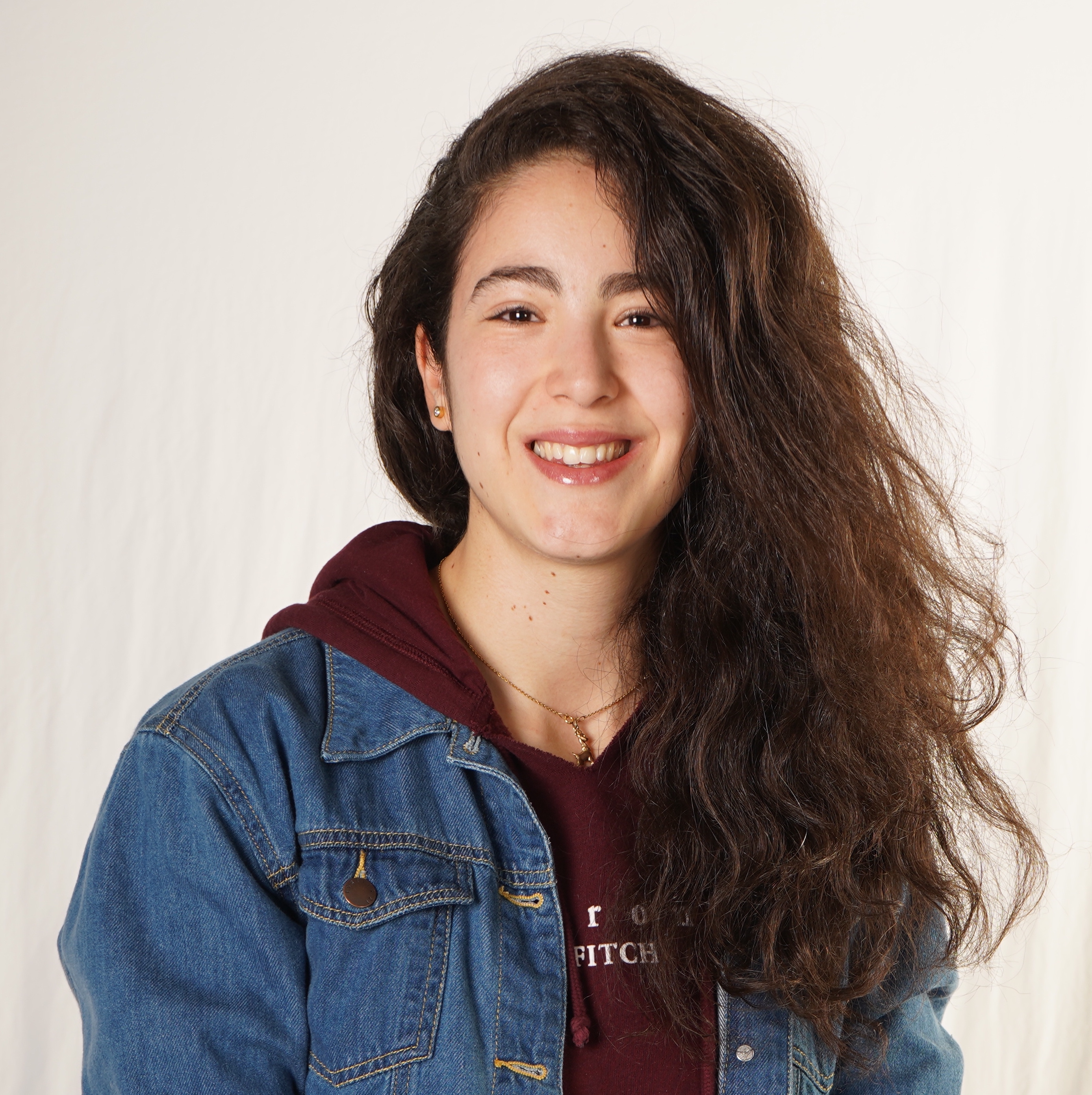}}]{Mariam Guizani} is a PhD student in Computer Science at Oregon State University. Her research area is in Human-Computer Interaction and Software Engineering. More specifically her research focuses on improving diversity in open source. She holds a Master of Science in Computer Science from Oregon State University. \end{IEEEbiography}
\vspace{-1.2cm}

\begin{IEEEbiography}[{\includegraphics[clip,keepaspectratio, width=1in,height=1.25in]{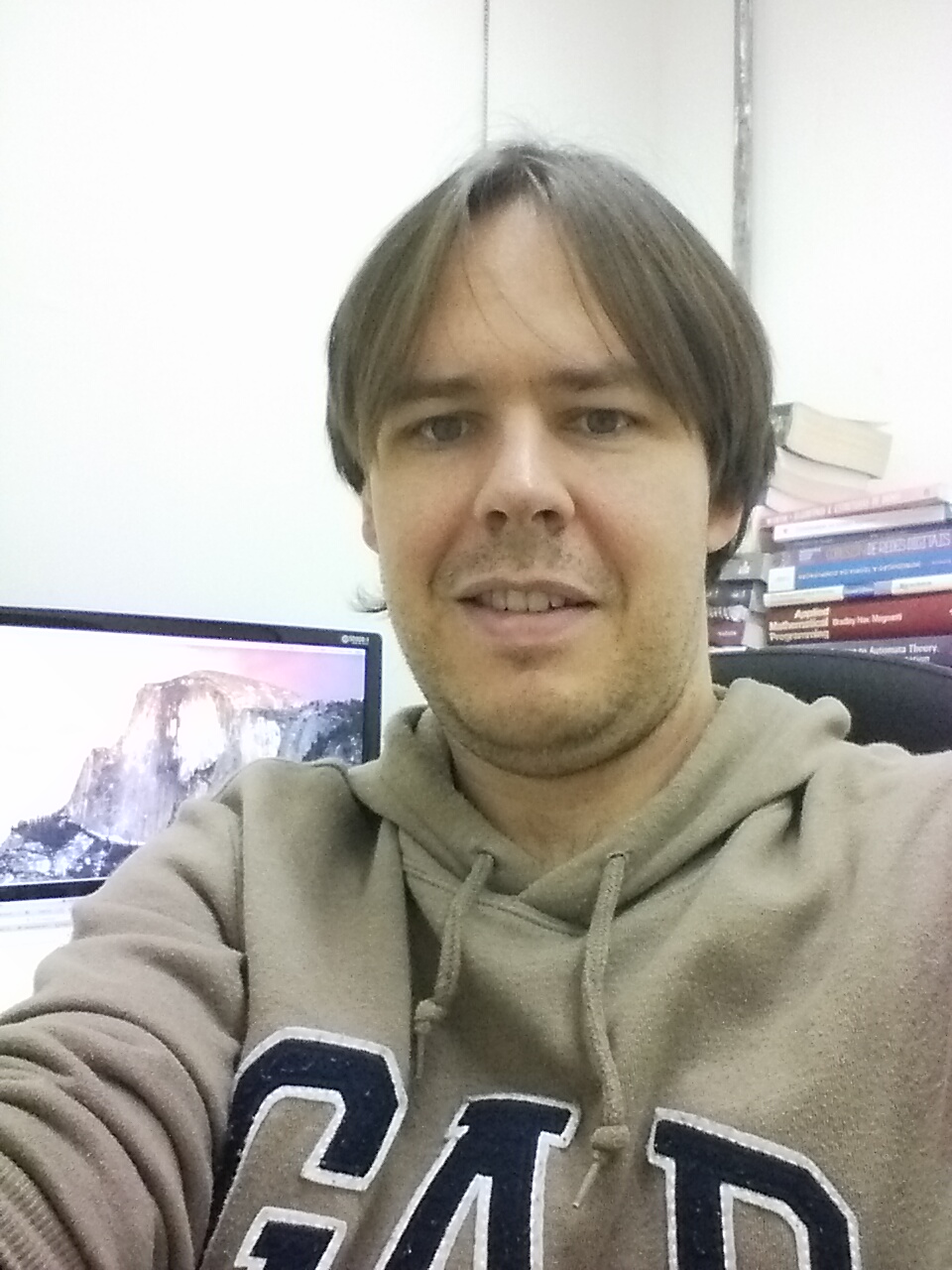}}\\]{Igor Wiese} is an Associate Professor in the Department of Computing at the Federal University of Technology – Parana, Brazil. He is interested in Mining Software Repositories, Human Aspects of Software Engineering, and related topics. Wiese holds a PhD degree in Computer Science from the University of São Paulo. More information is available at http://www.igorwiese.com.
\end{IEEEbiography}
\vspace{-1.2cm}

\begin{IEEEbiography}[{\includegraphics[width=1in,height=1.25in,keepaspectratio]{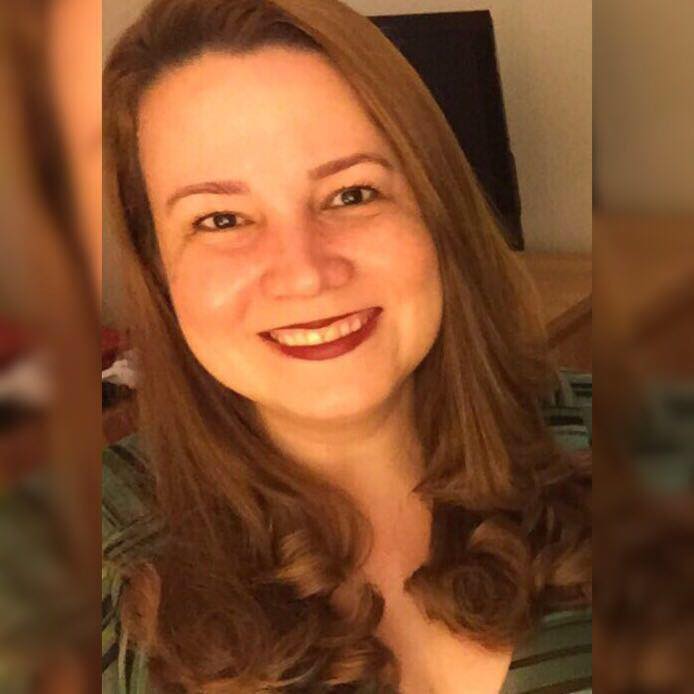}}]{Tayana Conte} holds a PhD in Systems Engineering and Computer from the Federal University of Rio de Janeiro (UFRJ). She is an associate professor at the Institute of Computing (IComp) of Federal University of Amazonas, heading the Usability and Software Engineering (USES) lab. Her research interests include the intersection between Software Engineering and Human-Computer Interaction, Software Quality, Human-Centered Computing, and Empirical Software Engineering. \end{IEEEbiography}
\vspace{-1.0cm}

\begin{IEEEbiography}[{\includegraphics[width=1in,height=1.25in,clip,keepaspectratio]{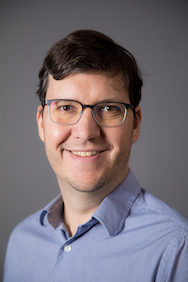}}]{Marco Gerosa} is an Associate Professor at the Northern Arizona University. He holds a PhD in Computer Science from the Pontifical Catholic University of Rio de Janeiro. He researches Software Engineering and CSCW. Recent projects include the development of strategies to support newcomers onboarding to open source communities and the design of chatbots for tourism. He published more than 200 papers and served on the program committee (PC) of important conferences, such as FSE, CSCW, SANER, and MSR.
\end{IEEEbiography}
\vspace{-1cm}
% trim={<left> <lower> <right> <upper>}

\begin{IEEEbiography}[{\includegraphics[width=1in,height=1.25in,clip,keepaspectratio]{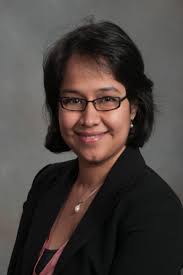}}]{Anita
Sarma} is an Associate Professor in the School of Electrical Engineering and Computer Science, at Oregon State University. She holds a PhD in Computer Science from the University of California, Irvine. Her research interests  intersect software engineering and human computer interaction, focusing on understanding and supporting end users and software developers. She has over 100 papers in journals and conferences. Her work has been recognized by an NSF CAREER award and several best paper awards.
\end{IEEEbiography}
\vspace{-1cm}

\begin{IEEEbiography}[{\includegraphics[width=1in,height=1.25in,clip,keepaspectratio]{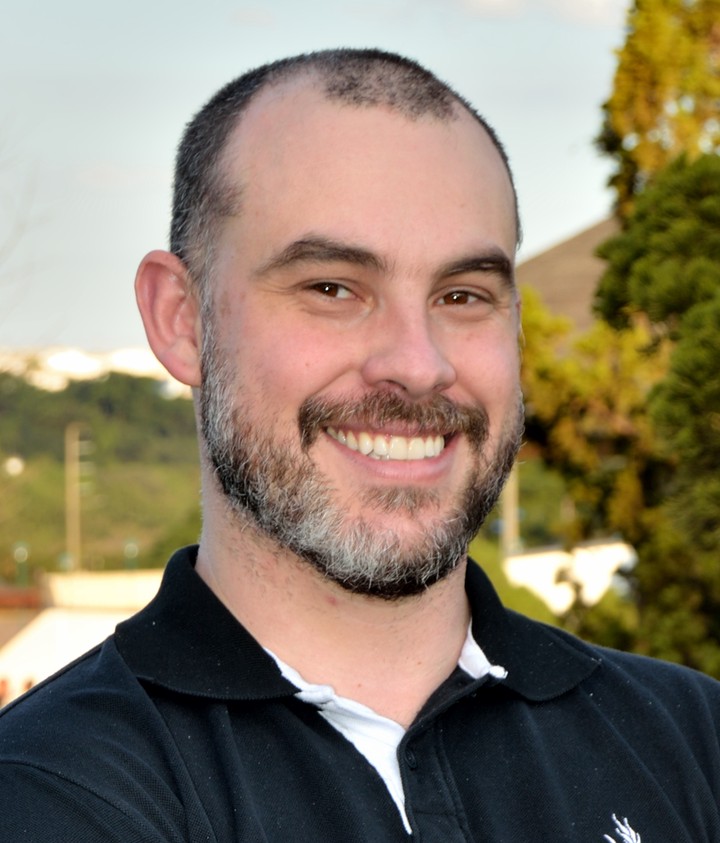}}]{Igor Steinmacher} is an Assistant Professor at the Federal University of Technology, Paraná. He received a PhD in computer science from the University of São Paulo. His topics of interest include human aspects of software engineering, behavior in open source software communities, mining software repositories, and software engineering education \& training.
\end{IEEEbiography}

\end{document}